\documentclass[preprint]{aastex63}

\usepackage{amsmath}
\usepackage{bm}

\received{}
\revised{}
\accepted{}
\submitjournal{}
\shorttitle{Constraints on Nuclear Saturation Properties}
\shortauthors{Choi et al.}

\begin{document}

\title{Constraints on Nuclear Saturation Properties from Terrestrial Experiments and Astrophysical Observations of Neutron Stars}

\correspondingauthor{Tsuyoshi Miyatsu}
\email{tsuyoshi.miyatsu@rs.tus.ac.jp}

\author{Soonchul Choi}
\affiliation{Department of Physics and OMEG institute, Soongsil University, Seoul 156-743, Republic of Korea}
\affiliation{Center for Exotic Nuclear Studies, Institute for Basic Science, Daejeon 34126, Republic of Korea}

\author[0000-0001-9186-8793]{Tsuyoshi Miyatsu}
\affiliation{Department of Physics, Faculty of Science and Technology, Tokyo University of Science, Noda, 278-8510, Japan}

\author[0000-0001-7810-5134]{Myung-Ki Cheoun}
\affiliation{Department of Physics and OMEG institute, Soongsil University, Seoul 156-743, Republic of Korea}

\author[0000-0002-8563-9262]{Koichi Saito}
\affiliation{Department of Physics, Faculty of Science and Technology, Tokyo University of Science, Noda, 278-8510, Japan}

\begin{abstract}
  Taking into account the terrestrial experiments and the recent astrophysical observations of neutron stars and gravitational-wave signals, we impose restrictions on the equation of state (EoS) for isospin-asymmetric nuclear matter.
  Using the relativistic mean-field model with SU(3) flavor symmetry, we investigate the impacts of effective nucleon mass, nuclear incompressibility, and slope parameter of nuclear symmetry energy on the nuclear and neutron-star properties.
  It is found that the astrophysical information of massive neutron stars and tidal deformabilities as well as the nuclear experimental data plays an important role to restrict the EoS for neutron stars.
  Especially, the softness of the nuclear EoS due to the existence of hyperons in the core gives stringent constraints on those physical quantities.
  Furthermore, it is possible to put limits on the curvature parameter of nuclear symmetry energy by means of the nuclear and astrophysical calculations.
\end{abstract}


\section{Introduction}
\label{sec:introduction}

Neutron stars are known to be the densest object in our universe.
Since the first discovery of a pulsar \citep{1965Natur.207...59H,Hewish:1968bj}, many theoretical studies regarding the understanding of neutron-star characteristics have been performed because the nature of neutron stars are broadly determined by the nuclear equation of state (EoS).
At the present, neutron stars are believed to be cosmological laboratories for dense nuclear matter \citep{Glendenning:1997wn,Lattimer:2006xb}.
Conversely, the accurate data from astrophysical observations can enable us to select the appropriate EoSs for neutron stars, and thus it may be possible to figure out the properties of nuclear matter around and beyond the nuclear saturation density.

Thanks to the advanced technological development in science, some invaluable information due to astrophysical observations have been reported in the last decade.
In particular, Shapiro delay measurements of a two-solar-mass neutron star have a great impact on the astrophysical and nuclear communities because it is very difficult to explain such massive neutron stars using the existing EoSs with exotic degrees of freedom, such as hyperons and/or kaon condensates \citep{Glendenning:1991es,SchaffnerBielich:2008kb}.
The binary millisecond pulsar, J1614-2230, has the mass of $1.97\pm0.04$ $M_{\sun}$ \citep{Demorest:2010bx}, and it is recently updated to $1.908\pm0.016$ $M_{\sun}$ \citep{Arzoumanian:2017puf}.
The mass of PSR J0348+0432 is also estimated to be $2.01\pm0.04$ $M_{\sun}$ by a combination of radio timing and precise spectroscopy of the white dwarf companion \citep{Antoniadis:2013pzd}.
Furthermore, an extremely massive millisecond pulsar, J0740+6620, has been found, and the mass is measured to be $2.14^{+0.10}_{-0.09}$ $M_{\sun}$ \citep{Cromartie:2019kug}.

In addition to the observations of massive neutron stars, a new type of observational data has been established by the gravitational wave (GW) from binary neutron star merger detected by the advanced LIGO and advanced Virgo observatories \citep{Abbott:2018exr,Abbott:2018wiz}.
Because the GW signals of binary neutron-star inspirals can potentially yield robust information on the nuclear EoS, it is quite useful to consider the tidal deformability of a neutron star \citep{Hinderer:2007mb,Hinderer:2009ca}.
The dimensionless tidal deformability, $\Lambda$, is defined as $\Lambda = \frac{2}{3}k_{2}\left(\frac{R}{M}\right)^{5}$,
where $k_{2}$ is the second Love number, and $M$ and $R$ are, respectively, the mass and radius of a neutron star.
Recently, many theoretical discussions have focused on the GW information and the tidal deformability \citep{Annala:2017llu,De:2018uhw,Most:2018hfd,Raithel:2018ncd,Lim:2018bkq,Capano:2019eae}.
They have reported that the tidal deformability of a canonical $1.4M_{\sun}$ neutron star, $\Lambda_{1.4}$, is highly sensitive to its radius, $R_{1.4}$, and the GW signals is of great use to give stringent constraints on the EoS for neutron-star matter \citep{Chatziioannou:2018vzf,Kim:2018aoi,Malik:2018zcf,Radice:2017lry,Tews:2018iwm,Tews:2019cap,Zhao:2018nyf,Lourenco:2018dvh,Wei:2018dyy}.

From the viewpoint of nuclear physics, the nuclear symmetry energy, $E_{\rm sym}$, is recognized to be a significant physical quantity to explain properties of finite nuclei and nuclear matter at low densities \citep{Li:2008gp,Danielewicz:2008cm}.
Although the characteristics of isospin-asymmetric nuclear matter in the density region beyond the nuclear saturation density are still under debate, much progress has been made in understanding the density dependence of $E_{\rm sym}$ based on various analyses of terrestrial experiments such as heavy-ion collisions \citep{Li:2014oda,Li:2019xxz,Zhang:2018vrx,Xie:2019sqb,Miyatsu:2020vzi}.
Recently many calculations have focused on the correlation between $E_{\rm sym}$ and $\Lambda$ to determine the EoSs for neutron-rich nuclear matter at supra-high densities \citep{Fattoyev:2017jql,Krastev:2018nwr,Raithel:2019ejc,Zhang:2018vbw}.

In the present study, we restrict the nuclear EoSs based on the recent data of terrestrial experiments and astrophysical observations using the relativistic mean-field (RMF) model with a nonlinear potential \citep{Walecka:1974qa,Boguta:1977xi,Serot:1984ey,ToddRutel:2005zz,Fattoyev:2010mx}.
We take into account the semi-empirical data deduced from the realistic $N$-$N$ interaction \citep{Katayama:2013zya,Sammarruca:2014zia} and the analyses of heavy-ion collisions \citep{Tsang:2012se,Russotto:2016ucm} in the intermediate density region.
Not only the tidal deformabilities \citep{Abbott:2018exr} but also the observed maximum masses of neutron stars \citep{Antoniadis:2013pzd,Cromartie:2019kug} are exploited to investigate the neutron-star properties.
Moreover, hyperons as well as nucleons are explicitly included in the core of a neutron star within SU(3) flavor symmetry \citep{Miyatsu:2011bc,Miyatsu:2013yta,Katayama:2012ge,Weissenborn:2011ut}, since the exotic degrees of freedom are known to soften the EoSs for neutron-star matter drastically and their effect is still an open question in the multi-messenger Era \citep{Kumar:2016dks,Li:2018ayl,Paschalidis:2017qmb,Zhou:2017pha,Zhu:2018ona,Li:2019tjx,Ribes:2019kno,Sahoo:2019qaq,Fortin:2020qin}.
At last, we present the calibrated EoSs for neutron stars and the preferable relations in nuclear saturation properties such as effective nucleon mass, nuclear incompressibility, and slope and curvature parameters of $E_{\rm sym}$.

This paper is organized as follows.
In Section \ref{sec:model}, a brief review of the RMF model in SU(3) flavor symmetry is presented.
Numerical results compatible for nuclear and neutron-star matter are presented with detailed discussions concerning the correlations among the important physical quantities in Section \ref{sec:result}.
Finally, we give a summary in Section \ref{sec:summary}.

\section{Theoretical framework}
\label{sec:model}

For describing the properties of nuclear and neutron-star matter, we employ the usual Lagrangian density in RMF approximation \citep{Walecka:1974qa,Serot:1984ey}.
In addition, for the purpose of studying the impact of strangeness in the core of a neutron star, not only the $\sigma$, $\omega$, and $\bm{\rho}$ mesons but also the strange mesons, namely the isoscalar, Lorentz scalar ($\sigma^{\ast}$) and vector ($\phi$) mesons, are taken into account in SU(3) flavor symmetry \citep{Miyatsu:2011bc,Miyatsu:2013yta,Katayama:2012ge,Weissenborn:2011ut}.
Since the charge neutrality and $\beta$ equilibrium conditions are imposed in neutron-star calculations, leptons must be introduced as well.
The Lagrangian density is thus chosen to be
\begin{align}
  \mathcal{L}
  & = \sum_{B}\bar{\psi}_{B}\left[i\gamma_{\mu}\partial^{\mu}
    - M_{B}^{\ast}(\sigma,\sigma^{\ast})
    - g_{\omega B}\gamma_{\mu}\omega^{\mu}
    - g_{\phi B}\gamma_{\mu}\phi^{\mu}
    - g_{\rho B}\gamma_{\mu}\bm{\rho}^{\mu}\cdot\bm{I}_{B}\right]\psi_{B}
    \nonumber \\
  & + \frac{1}{2}\left(\partial_{\mu}\sigma\partial^{\mu}\sigma-m_{\sigma}^{2}\sigma^{2}\right)
    + \frac{1}{2}\left(\partial_{\mu}\sigma^{\ast}\partial^{\mu}\sigma^{\ast}-m_{\sigma^{\ast}}^{2}\sigma^{\ast2}\right)
    \nonumber \\
  & + \frac{1}{2}m_{\omega}^{2}\omega_{\mu}\omega^{\mu}-\frac{1}{4}W_{\mu\nu}W^{\mu\nu}
    + \frac{1}{2}m_{\phi}^{2}\phi_{\mu}\phi^{\mu}-\frac{1}{4}P_{\mu\nu}P^{\mu\nu}
    + \frac{1}{2}m_{\rho}^{2}\bm{\rho}_{\mu}\cdot\bm{\rho}^{\mu}-\frac{1}{4}\bm{R}_{\mu\nu}\cdot\bm{R}^{\mu\nu}
    \nonumber \\
  & - U_{\rm NL}(\sigma,\omega^{\mu},\bm{\rho}^{\mu})
    + \sum_{\ell}\bar{\psi}_{\ell}\left(i\gamma_{\mu}\partial^{\mu}-m_{\ell}\right)\psi_{\ell},
    \label{eq:Lagrangian}
\end{align}
where $W_{\mu\nu}=\partial_{\mu}\omega_{\nu}-\partial_{\nu}\omega_{\mu}$, $P_{\mu\nu}=\partial_{\mu}\phi_{\nu}-\partial_{\nu}\phi_{\mu}$, and $\bm{R}_{\mu\nu}=\partial_{\mu}\bm{\rho}_{\nu}-\partial_{\nu}\bm{\rho}_{\mu}$ with $\psi_{B (\ell)}$ being the baryon (lepton) field, $\bm{I}_B$ being the isospin matrix for baryon, and $m_{\ell}$ being the lepton mass.
The sum $B$ runs over the octet baryons, $N$ (proton and neutron), $\Lambda$, $\Sigma^{+,0,-}$, and $\Xi^{0,-}$, and the sum $\ell$ is for the leptons, $e^{-}$ and $\mu^{-}$.
The $\omega$-, $\phi$-, and $\rho$-$B$ coupling constants are respectively denoted by $g_{\omega B}$, $g_{\phi B}$, and $g_{\rho B}$.
The effective baryon mass, $M_{B}^{\ast}$, in matter is simply expressed as $ M_{B}^{\ast}(\sigma,\sigma^{\ast})=M_{B}-g_{\sigma B}\sigma-g_{\sigma^{\ast}B}\sigma^{\ast}$ with $M_{B}$ being the mass in vacuum, and $g_{\sigma B}$ ($g_{\sigma^{\ast}B}$) being the $\sigma$-$B$ ($\sigma^{\ast}$-$B$) coupling constant.
Additionally, to obtain a quantitative description of nuclear ground-state properties, the minimum set of a nonlinear potential \citep{Boguta:1977xi,ToddRutel:2005fa,Miyatsu:2013yta,Hornick:2018kfi},
\begin{equation}
  U_{NL}(\sigma,\omega^{\mu},\bm{\rho}^{\mu})
  = \frac{1}{3}g_{2}\sigma^{3} + \frac{1}{4}g_{3}\sigma^{4}
  - \Lambda_{\omega\rho}(\omega_{\mu}\omega^{\mu})(\bm{\rho}_{\mu}\cdot\bm{\rho}^{\mu}),
  \label{eq:NL-potential}
\end{equation}
is introduced in Equation \eqref{eq:Lagrangian}.
Here, the potential involves three coupling constants, $g_{2}$, $g_{3}$, and $\Lambda_{\omega\rho}$.
In the present study, the hadron, meson, and lepton masses are taken as follows: $M_{N}=939$ MeV, $M_{\Lambda}=1116$ MeV, $M_{\Sigma}=1193$ MeV, $M_{\Xi}=1318$ MeV, $m_{\sigma}=500$ MeV, $m_{\omega}=783$ MeV, $m_{\rho}=770$ MeV, $m_{\sigma^{\ast}}=975$ MeV, $m_{\phi}=1020$ MeV, $m_{e}=0.511$ MeV, and $m_{\mu}=105.7$ MeV.

In RMF approximation, the meson fields are replaced by the constant mean-field values: $\bar{\sigma}$, $\bar{\omega}$, $\bar{\sigma}^{\ast}$, $\bar{\phi}$ and $\bar{\rho}$ (the $\rho^{0}$ field).
The equations of motion for the baryon and meson fields in uniform matter are thus given by
\begin{equation}
  \left[i\gamma_{\mu}\partial^{\mu}
    - M_{B}^{\ast}(\bar{\sigma},\bar{\sigma}^{\ast})
    - g_{\omega B}\gamma_{0}\bar{\omega}
    - g_{\phi B}\gamma_{0}\bar{\phi}
    - g_{\rho B}\gamma_{0}(\bm{I}_{B})_{3}\bar{\rho}\right]\Psi_{B}
  = 0,
  \label{eq:Dirac}
\end{equation}
\begin{align}
  m_{\sigma}^{2}\bar{\sigma} + g_{2}\bar{\sigma}^{2} + g_{3}\bar{\sigma}^{3}
  & = \sum_{B} g_{\sigma B}\rho_{B}^{s},
    \label{eq:EoM-sigma} \\
  m_{\sigma^{\ast}}^{2}\bar{\sigma}^{\ast}
  & = \sum_{B} g_{\sigma^{\ast} B}\rho_{B}^{s},
    \label{eq:EoM-sigma-star} \\
  \left(m_{\omega}^{2}+2\Lambda_{\omega\rho}\bar{\rho}^{2}\right)\bar{\omega}
  & = \sum_{B} g_{\omega B} \rho_{B},
    \label{eq:EoM-omega} \\
  m_{\phi}^{2}\bar{\phi}
  & = \sum_{B} g_{\phi B} \rho_{B},
    \label{eq:EoM-phi} \\
  \left(m_{\rho}^{2}+2\Lambda_{\omega\rho}\bar{\omega}^{2}\right)\bar{\rho}
  & = \sum_{B} g_{\rho B}(\bm{I}_{B})_{3}\rho_{B},
    \label{eq:EoM-rho}
\end{align}
where the scalar density, $\rho_{B}^{s}$, and the baryon density, $\rho_{B}$, read
\begin{equation}
  \rho_{B}^{s}
  = \frac{1}{\pi^{2}}\int_{0}^{k_{F_{B}}}dk~k^{2}\frac{M_{B}^{\ast}(\bar{\sigma},\bar{\sigma}^{\ast})}
  {\sqrt{k^{2}+M_{B}^{\ast2}(\bar{\sigma},\bar{\sigma}^{\ast})}}, \quad
  \rho_{B}
  = \frac{k^{3}_{F_{B}}}{3\pi^{2}},
  \label{eq:baryon-density}
\end{equation}
with $k_{F_B}$ being the Fermi momentum for baryon $B$.

With the self-consistent calculations of the meson fields given in Equations \eqref{eq:EoM-sigma}--\eqref{eq:EoM-rho}, the total energy density, $\varepsilon$, and pressure, $P$, in neutron-star matter are given by
\begin{align}
  \varepsilon
  & = \sum_{B}\frac{1}{\pi^{2}}\int_{0}^{k_{F_{B}}}dk~k^{2}\sqrt{k^{2}+M_{B}^{\ast2}(\bar{\sigma},\bar{\sigma}^{\ast})}
    + \sum_{\ell}\frac{1}{\pi^{2}}\int_{0}^{k_{F_{\ell}}}dk~k^{2}\sqrt{k^{2}+m_{\ell}^{2}},
    \nonumber \\
  & + \frac{1}{2}m_{\sigma}^{2}\bar{\sigma}^{2}
    + \frac{1}{3}g_{2}\bar{\sigma}^{3}
    + \frac{1}{4}g_{3}\bar{\sigma}^{4}
    + \frac{1}{2}m_{\sigma^{\ast}}^{2}\bar{\sigma}^{\ast2}
    + \frac{1}{2}m_{\omega}^{2}\bar{\omega}^{2}
    + \frac{1}{2}m_{\phi}^{2}\bar{\phi}^{2}
    + \frac{1}{2}m_{\rho}^{2}\bar{\rho}^{2}
    + 3\Lambda_{\omega\rho} \bar{\omega}^{2} \bar{\rho}^{2},
    \label{eq:engy-density} \\
  P
  & = \frac{1}{3}\sum_{B}\frac{1}{\pi^{2}}\int_{0}^{k_{F_{B}}}dk~
    \ \frac{k^{4}}{\sqrt{k^{2}+M_{B}^{\ast2}(\bar{\sigma},\bar{\sigma}^{\ast})}}
    + \frac{1}{3}\sum_{\ell}\frac{1}{\pi^{2}}\int_{0}^{k_{F_{\ell}}}dk~\frac{k^{4}}{\sqrt{k^{2}+m_{\ell}^{2}}}
    \nonumber \\
  & - \frac{1}{2}m_{\sigma}^{2}\bar{\sigma}^{2}
    - \frac{1}{3}g_{2}\bar{\sigma}^{3}
    - \frac{1}{4}g_{3}\bar{\sigma}^{4}
    - \frac{1}{2}m_{\sigma^{\ast}}^{2}\bar{\sigma}^{\ast2}
    + \frac{1}{2}m_{\omega}^{2}\bar{\omega}^{2}
    + \frac{1}{2}m_{\phi}^{2}\bar{\phi}^{2}
    + \frac{1}{2}m_{\rho}^{2}\bar{\rho}^{2}
    + \Lambda_{\omega\rho} \bar{\omega}^{2} \bar{\rho}^{2}.
    \label{eq:pressure}
\end{align}

\section{Numerical results and discussions}
\label{sec:result}

We here present how to calibrate the nuclear EoS using the RMF model with a nonlinear potential.
First, the coupling constants for nucleon ($N$) are determined so as to reproduce the saturation properties of nuclear matter from terrestrial experiments as well as theoretical calculations.
And then, the calibrated EoSs are applied to the calculations of neutron-star matter, compared with the recent data of astrophysical observations, such as the maximum mass of a neutron star, $M_{\rm max}$, and the tidal deformability of a canonical $1.4M_{\sun}$ neutron star, $\Lambda_{1.4}$.
Especially, we focus on the effect of hyperons ($Y$) in the core of a neutron star.
Finally, we give some constraints on the nuclear saturation properties, which are not known well, so as to satisfy the data based on both nuclear experiments and astrophysical observations.

In order to deal with the properties of nuclear matter, it is very useful to consider the expansion of isospin asymmetric EoS with a power series in the isospin asymmetry, $\delta=(\rho_{n}-\rho_{p})/n_{B}$, where the total baryon density is defined as $n_{B}=\sum_{B}\rho_{B}$ \citep{Chen:2009wv}.
The binding energy per nucleon is generally written as $E(n_{B},\delta)=E_{0}(n_{B})+E_{\rm sym}(n_{B})\delta^{2}+\mathcal{O}(\delta^{4})$, where $E_{0}(n_{B})$ is the binding energy of symmetric nuclear matter and $E_{\rm sym}(n_{B})$ is the nuclear symmetry energy,
\begin{equation}
  E_{\rm sym}(n_{B}) = \left.\frac{1}{2!}\frac{\partial^{2}E(n_{B},\delta)}{\partial\delta^{2}}\right|_{\delta=0}.
  \label{eq:Esym-def}
\end{equation}
Besides, $E_{0}(n_{B})$ and $E_{\rm sym}(n_{B})$ can be expanded around the nuclear saturation density, $n_{0}$, as
\begin{align}
  E_{0}(n_{B})
  & = E_{0}(n_{0}) + \frac{K_{0}}{2!}\chi^{2} + \mathcal{O}(\chi^{3}), \\
  E_{\rm sym}(n_{B})
  & = E_{\rm sym}(n_{0}) + L\chi + \frac{K_{\rm sym}}{2!}\chi^{2} + \mathcal{O}(\chi^{3}),
\end{align}
with $\chi=(n_{B}-n_{0})/3n_{0}$ being a dimensionless variable characterizing the deviations of $n_{B}$ from $n_{0}$.
The incompressibility coefficient of symmetric nuclear matter, $K_{0}$, and the slope and curvature parameters of nuclear symmetry energy, $L$ and $K_{\rm sym}$, are respectively given by
\begin{equation}
  K_{0} = \left.9n_{0}^{2}\frac{d^{2}E_{0}(n_{B})}{dn_{B}^{2}}\right|_{n_{B}=n_{0}}, \quad
  L = \left.3n_{0}\frac{dE_{\rm sym}(n_{B})}{dn_{B}}\right|_{n_{B}=n_{0}}, \quad
  K_{\rm sym} = \left.9n_{0}^{2}\frac{d^{2}E_{\rm sym}(n_{B})}{dn_{B}^{2}}\right|_{n_{B}=n_{0}}.
\end{equation}

\subsection{Terrestrial experiments and theoretical calculations}
\label{subsec:nuclear-constraint}

The nuclear saturation properties have been extensively studied so far, and the binding energy per nucleon and nuclear symmetry energy at the saturation density, $E_{0}(n_{0})$ and $E_{\rm sym}(n_{0})$, are determined with a highly accurate precision \citep{Dutra:2012mb,Dutra:2014qga,Baldo:2016jhp}.
However, the effective nucleon mass, $M_{N}^{\ast}$ and the higher-order physical quantities, e.g.~the incompressibility coefficient, $K_{0}$, and the slope and curvature parameters, $L$ and $K_{\rm sym}$, still have a large ambiguity even at $n_{0}$ \citep{Li:2013ola,Zhang:2018vrx}.
As for the quantities, the recent standard values are hence employed as follows: we set $E_{0}=-16.0$ MeV and $E_{\rm sym}=32.0$ MeV at $n_{0}=0.16$ fm$^{-3}$, but $M_{N}^{\ast}$, $K_{0}$, and $L$ are supposed to be varied in the range of $0.50\leq M_{N}^{\ast}/M_{N}\leq0.80$, $180\leq K_{0}$ (MeV) $\leq320$, and $30\leq L$ (MeV) $\leq100$, respectively.

The determining procedure of the coupling constants in the present study follows that in \cite{Miyatsu:2013yta}: Using SU(3) flavor symmetry in the couplings of vector mesons to nucleons, the isoscalar coupling constants, $g_{\sigma N}$, $g_{\omega N}$, $g_{2}$, and $g_{3}$, are determined so as to reproduce $E_{0}$, $P$, $M_{N}^{\ast}$ and $K_{0}$ at $n_{0}$ with the assumption of $g_{\sigma^{\ast}N}=0$.
The values of a mixing angle, $\theta_{v}$, and a ratio of the octet to singlet couplings, $z$, are, respectively, chosen to be $\theta_{v}=37.50^{\circ}$ and $z=0.1949$, which are the suggested values in the Nijmegen extended-soft-core (ESC) model \citep{Rijken:2010zzb}.
Moreover, the coupling constants related to the isovector mesons, $g_{\rho N}$ and $\Lambda_{\omega\rho}$, are fixed to duplicate $E_{\rm sym}$ and $L$ at $n_{0}$.

\begin{figure}[ht!]
  \plotone{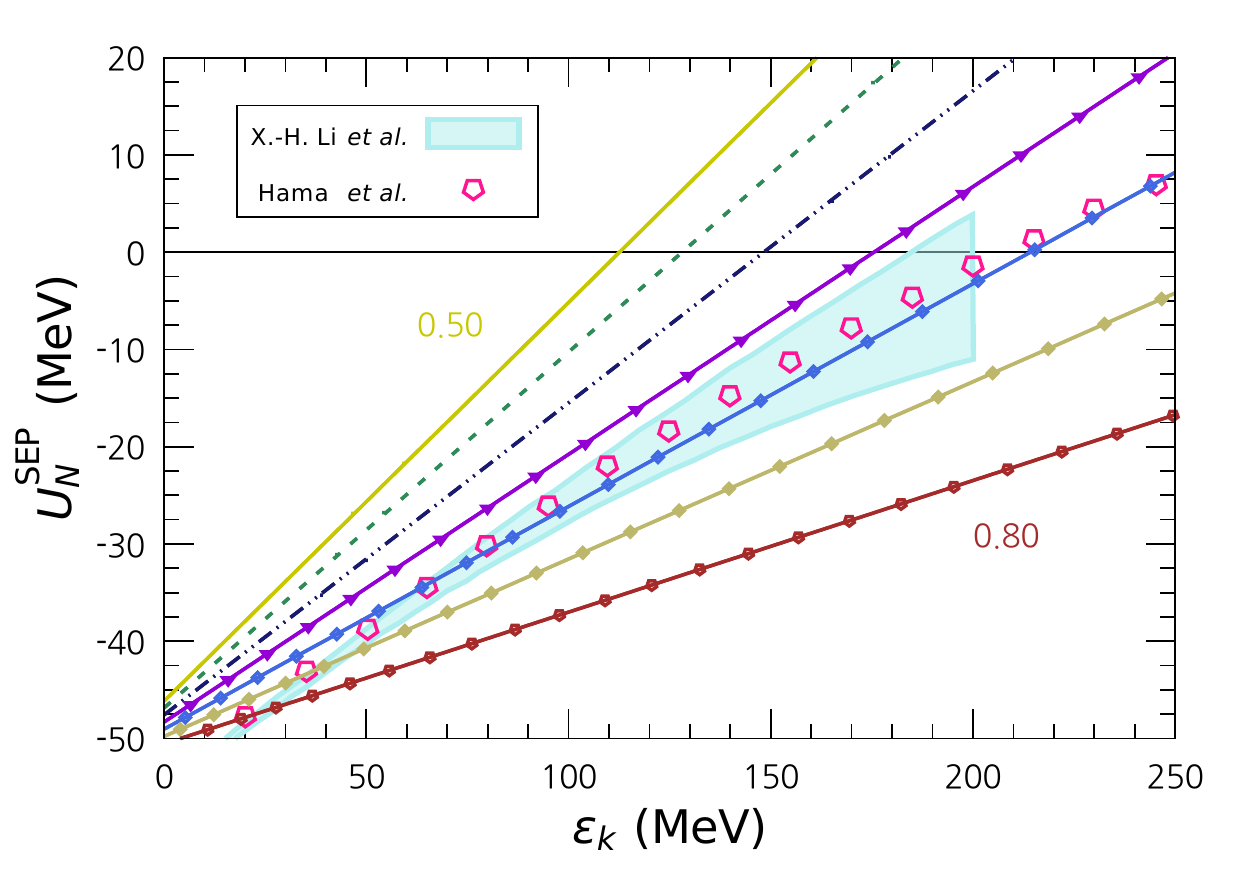}
  \caption{
    Energy dependence of single-nucleon potential, $U_{N}^{\rm SEP}$, in symmetric nuclear matter at $n_{0}$.
    We show the results for $M_{N}^{\ast}/M_{N}=0.5$--$0.8$ by a $0.05$ step.
    For details, see the text.
    \label{fig:1}}
\end{figure}
In order to study a limit on $M_{N}^{\ast}$ from the experiments and theoretical calculations, we consider the single-nucleon potential, $U_{N}^{\rm SEP}$, based on the so-called Schr\"{o}dinger-equivalent potential (SEP) \citep{Jaminon:1981xg,Chen:2007ih}:
\begin{equation}
  U_{N}^{\rm SEP}(k,\varepsilon_{k})
  = \Sigma_{N}^{s}(k)
  - \frac{E_{N}(k)}{M_{N}}\Sigma_{N}^{0}(k)
  + \frac{1}{2M_{N}}\left(\left[\Sigma_{N}^{s}(k)\right]^{2}- \left[\Sigma_{N}^{0}(k)\right]^{2}\right),
  \label{eq:USEP}
\end{equation}
where the nucleon kinetic energy, $\varepsilon_{k}$, reads $\varepsilon_{k}=E_{N}-M_{N}$ with $E_{N}$ being the single-particle energy.
The Lorentz-covariant scalar and vector self-energies for nucleon are respectively given by $\Sigma^{s}_{N}=-g_{\sigma}\bar{\sigma}-g_{\sigma^{\ast}}\bar{\sigma}^{\ast}$ and $\Sigma^{0}_{N}=-g_{\omega}\bar{\omega}-g_{\phi}\bar{\phi}-g_{\rho}(\bm{I}_{N})_{3}\bar{\rho}$ with the meson fields in Equations \eqref{eq:EoM-sigma}--\eqref{eq:EoM-rho}.
We show $U_{N}^{\rm SEP}$ in symmetric nuclear matter at $n_{0}$ with some experimental data in Figure \ref{fig:1}.
The shaded band reveals the results of nucleon-optical-model potential extracted from analyzing the nucleon-nucleus scattering data, denoted by X.-H.~Li et al. \citep{Li:2013ck}.
We also include the results of $U_{N}^{\rm SEP}$ obtained by the Dirac phenomenology for elastic proton-nucleus scattering data calculated by Hama {\it et al.} \citep{Hama:1990vr}.
It is found that the satisfied values of $M_{N}^{\ast}$ is roughly estimated to be $0.65\leq M_{N}^{\ast}/M_{N}\leq0.75$ in the current energy region.
We note that the case of $M_{N}^{\ast}/M_{N}=0.69$ is the optimum condition for satisfying the scattering data.

\begin{figure*}[ht!]
  \plottwo{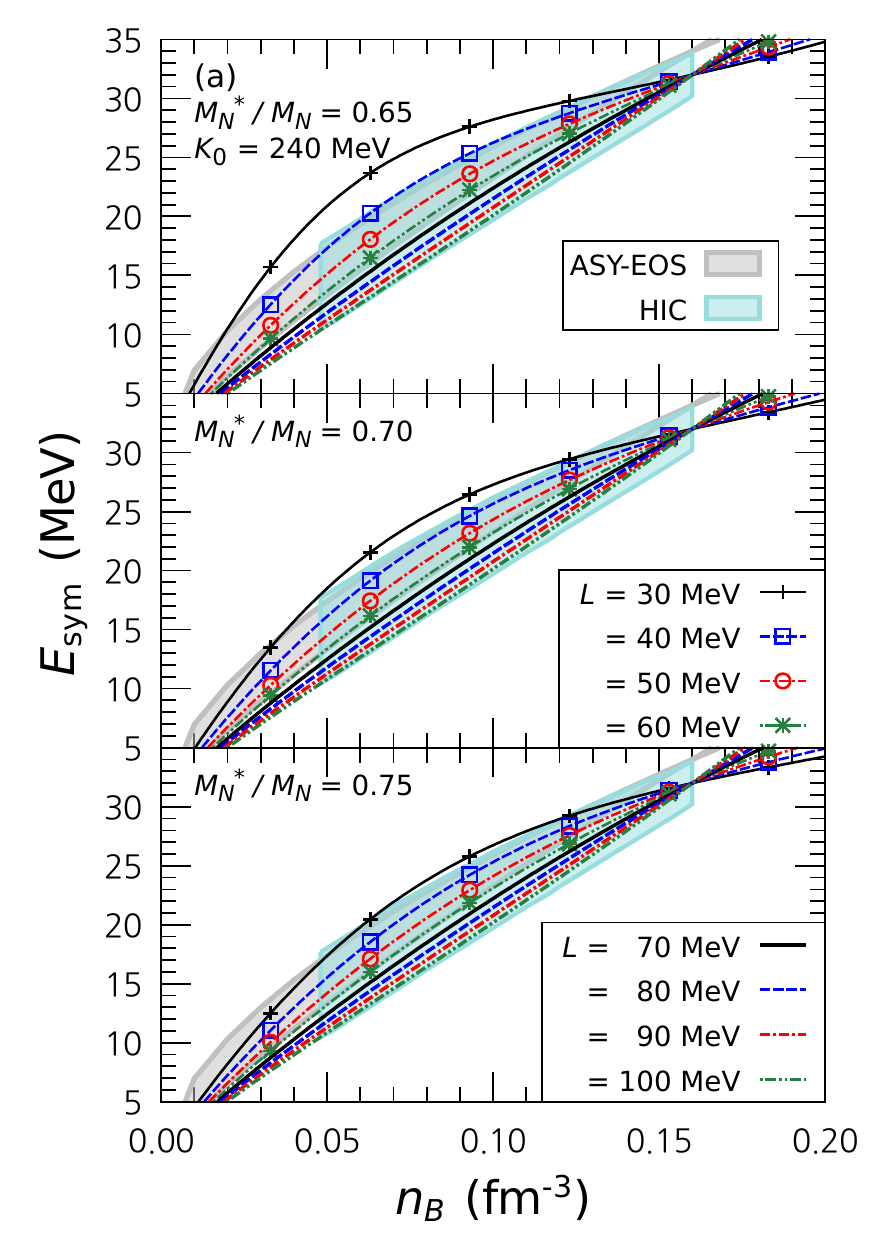}{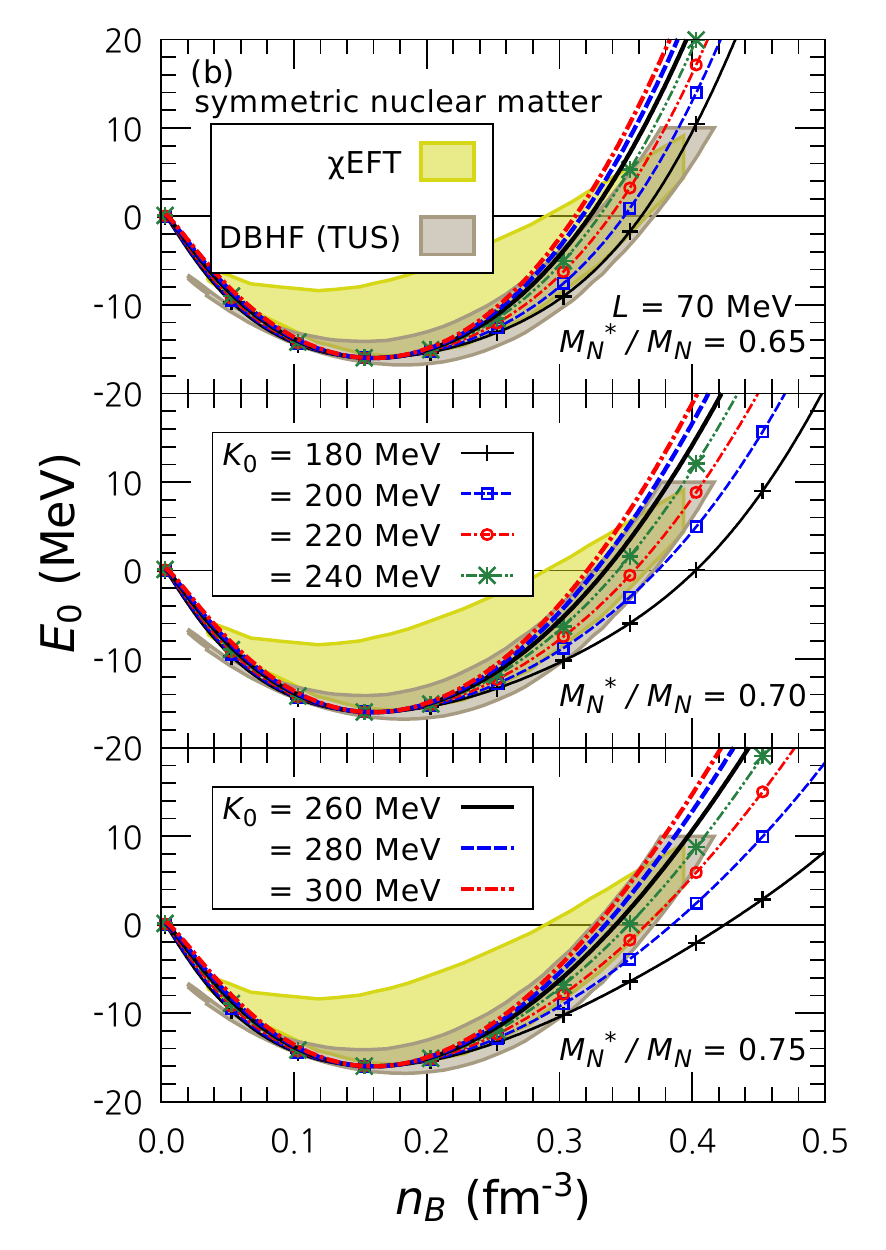}
  \caption{
    (a) Nuclear symmetry energy, $E_{\rm sym}$, and (b) binding energy per nucleon, $E_{0}$, as a function of $n_{B}$.
    In both upper (middle) [bottom] panels, we give the results in the case of $M_{N}^{\ast}/M_{N}=$ 0.65 (0.70) [0.75].
    The shaded bands in Figure (a) show the experimental data based on heavy-ion collisions, indicated by HIC and ASY-EOS \citep{Tsang:2012se,Russotto:2016ucm}.
    In Figure (b), we also present the theoretical results based on realistic $N$-$N$ interactions using the Dirac-Brueckner-Hartree-Fock (DBHF) calculation \citep{Katayama:2013zya} or the chiral effective field theory ($\chi$EFT) \citep{Sammarruca:2014zia}.
    \label{fig:2}}
\end{figure*}
In Figure \ref{fig:2}, the density dependence of $E_{\rm sym}$ and $E_{0}$ are depicted in the three cases of $M_{N}^{\ast}/M_N=0.65$, $0.70$, and $0.75$, which are deduced from the restriction of $M_{N}^{\ast}$ shown in Figure \ref{fig:1}.
In Figure \ref{fig:2}(a), we also present the experimental results obtained from heavy-ion collisions \citep{Tsang:2012se,Russotto:2016ucm}.
It is found that $E_{\rm sym}$ strongly depends on $L$ below $n_{0}$, and $L$ should be larger than 40 MeV to match the experimental data in all the cases.
Because of the less dependence of $K_{0}$, we only show the results in the case of $K_{0}=240$ MeV.

The $E_{0}$ in the intermediate density region is presented in Figure \ref{fig:2}(b), where $K_{0}$ varies from 180 MeV to 300 MeV.
Although the difference among the results with various $K_{0}$ is not so large below $n_0$, $E_{0}$ is sensitive to $K_{0}$ and $M_{N}^{\ast}$ above $n_{0}$.
On the other hand, since $L$ has little influence on $E_{0}$, we only show the results in the case of $L=70$ MeV, which is a middle value in the acceptable range of $L$ explained in Figure \ref{fig:2}(a).
Compared with the realistic calculations in the Dirac-Brueckner-Hartree-Fock (DBHF) theory \citep{Katayama:2013zya} and the chiral effective field theory ($\chi$EFT) \citep{Sammarruca:2014zia}, it is possible to impose some constraints on the nuclear EoSs at higher densities.
We find that the allowed range of $K_{0}$ can be restricted to be $180\leq K_{0}$ (MeV) $\leq230$, $210\leq K_{0}$ (MeV) $\leq270$, and $235\leq K_{0}$ (MeV) $\leq305$ for $M_{N}^{\ast}/M_{N}=0.65$, $0.70$, and $0.75$, respectively.

\subsection{Astrophysical observations of neutron stars}
\label{subsec:star-constraint}

The characteristics of a neutron star are, in general, estimated by solving the Tolman-Oppenheimer-Volkoff (TOV) equation with the EoS for neutron-star matter in which the charge neutrality and $\beta$ equilibrium under weak processes are imposed \citep{Tolman:1934za,Oppenheimer:1939ne}.
Since the radius of a neutron star is remarkably sensitive to the EoS at very low densities, we adopt the EoS for nonuniform matter below $n_{B}=0.068$ fm$^{-3}$, where nuclei are taken into account using the Thomas-Fermi calculation \citep{Miyatsu:2013hea}.
Moreover, the coupling constants for hyperons are determined so as to fit the experimental data of hypernuclei and the Nagara event in SU(3) flavor symmetry: $U_{\Lambda}^{(N)}=-28$ MeV, $U_{\Sigma}^{(N)}=+30$ MeV, $U_{\Xi}^{(N)}=-18$ MeV, and $U_{\Lambda}^{(\Lambda)}\simeq-5$ MeV, with $U_{Y}^{(j)}$ being the potential depth for $Y$ in matter of the baryon species $j$ \citep{Schaffner:1995th,Takahashi:2001nm,Yang:2008am,Miyatsu:2013yta}.
As for a potential depth for $Y$, we employ $U_{Y}^{\rm SEP}$ given in Equation \eqref{eq:USEP}.

\begin{figure*}[t!]
  \plotone{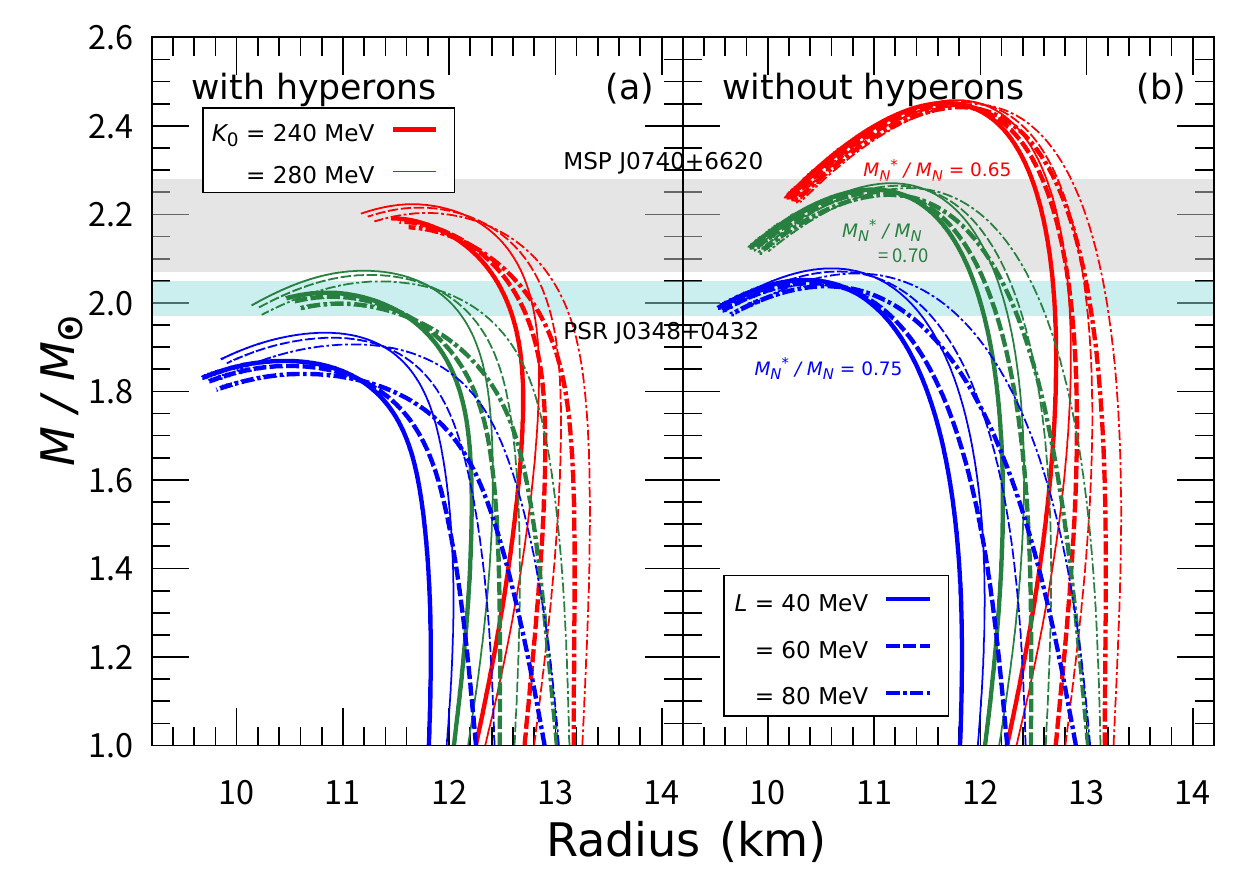}
  \caption{Mass-radius relations of a neutron star (a) with or (b) without hyperons in the various combinations of $M_{N}^{\ast}/M_{N}$, $L$, and $K_{0}$.
    In both panels, the red (green) [blue] line corresponds to the case of $M_N^\ast/M_N^{}=0.65$ ($0.70$) [$0.75$], the solid (dashed) [dotted-dashed] line is for $L=40$ ($60$) [$80$] MeV, and the case of $K_{0}=240$ ($280$) MeV is given by the thick (thin) line.
    The shaded bands also show the observation data of PSR J0348+0432 ($2.01\pm0.04$ $M_{\sun}$) and MSP J0740+6620 ($2.14^{+0.10}_{-0.09}$ $M_{\sun}$) \citep{Antoniadis:2013pzd,Cromartie:2019kug}.
    \label{fig:3}}
\end{figure*}
The mass-radius relations of a neutron star with and without hyperons are presented in Figure \ref{fig:3}.
As is well known, if hyperons are taken into account in the core, the maximum mass of a neutron star, $M_{\rm max}$, is reduced drastically.
Thus, in order to explain the observed masses of heavy neutron stars \citep{Antoniadis:2013pzd,Cromartie:2019kug}, it is possible to give more severe constraints on the neutron-star EoSs by considering hyperons.
It is found that $M_{\rm max}$ is very sensitive to $M_{N}^{\ast}$ at $n_{0}$, and the EoSs for $M_{N}^{\ast}/M_{N}>0.70$ are ruled out by the observation data once hyperons are included.
If we combine both restrictions of $M_{\rm max}$ (with hyperons) and $U_{N}^{\rm SEP}$, shown in Figure \ref{fig:1}, we can estimate the more refined condition of $M_{N}^{\ast}$ to be $0.65\leq M_{N}^{\ast}/M_{N}\leq0.70$ at $n_{0}$.
Meanwhile, $L$ and $K_{0}$ have a small impact on $M_{\rm max}$, and they rather affect the neutron-star radius.

\begin{figure*}[ht!]
  \plotone{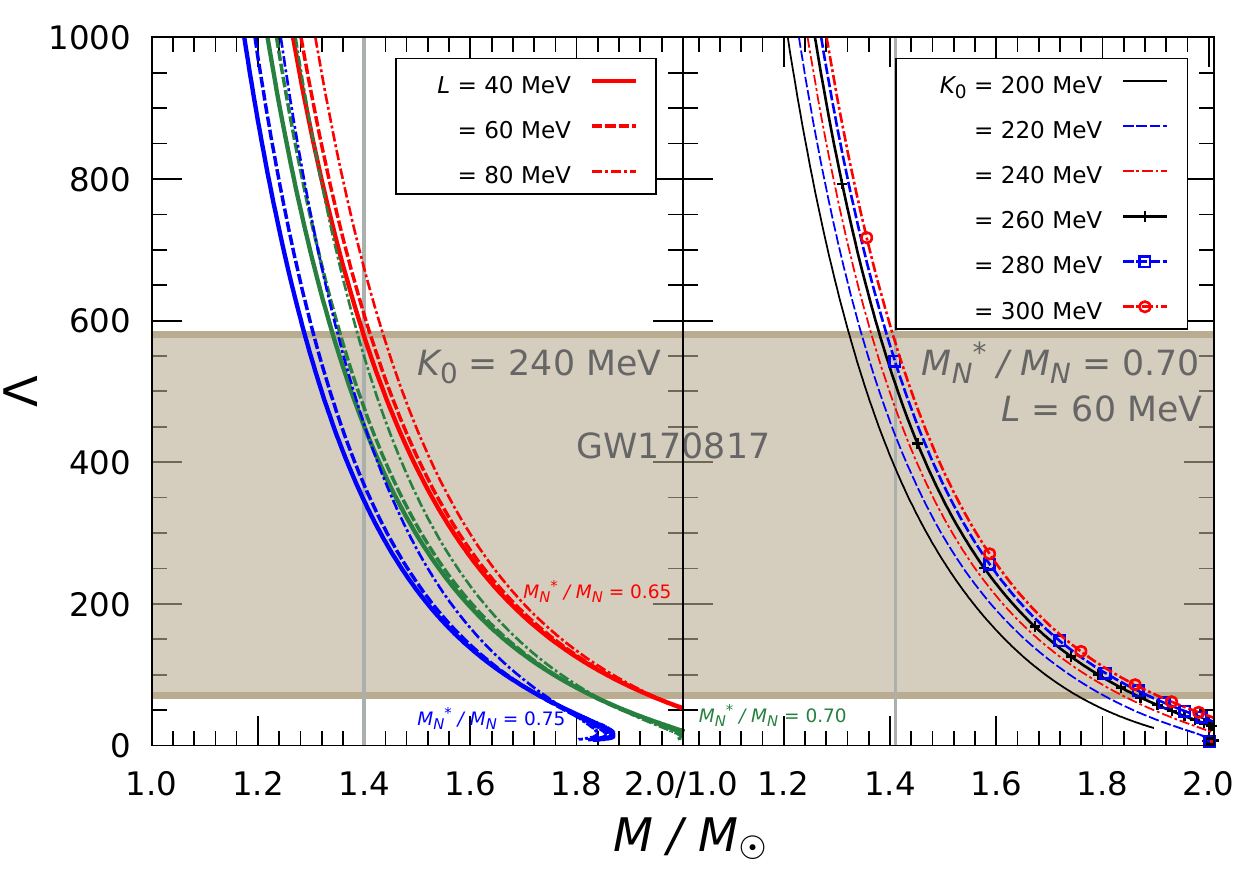}
  \caption{Tidal deformability of a neutron star with hyperons, $\Lambda$, as a function of $M/M_{\sun}$.
    The panel (a) shows the influence of $M_{N}^{\ast}/M_{N}$ and $L$ for $K_{0}=240$ MeV, and the panel (b) expresses the dependence of $K_{0}$ for $M_{N}^{\ast}/M_{N}=0.70$ and $L=60$ MeV.
    The shaded band also shows the astrophysical constraint on the tidal deformability of a canonical $1.4M_{\sun}$ neutron star from the merger event, GW170817 ($70<\Lambda_{1.4}<580$) \citep{Abbott:2018exr}.
    \label{fig:4}}
\end{figure*}
In Figure \ref{fig:4}, the dimensionless tidal deformability of a neutron star with hyperons, $\Lambda$,  is presented with the astrophysical constraint on the GW signals from binary neutron star merger, GW170817, detected by the advanced LIGO and advanced Virgo observatories \citep{Abbott:2018exr,Abbott:2018wiz}.
Hereafter, hyperons are taken into account in all the calculations.
With the observation data, it is possible to impose constraints on the EoSs for neutron stars using the tidal deformability of a canonical $1.4M_{\sun}$ neutron star, $\Lambda_{1.4}$.
It is found that $\Lambda$ becomes small as $M_{N}^{\ast}$ at $n_{0}$ increases, but the larger $L$, which gives the larger radius of a neutron star shown in Figure \ref{fig:3}, consequently brings the larger $\Lambda$ as shown in Figure \ref{fig:4}(a).
We also see that, in Figure \ref{fig:4}(b), there is a strong correlation between $K_{0}$ and $\Lambda$, and the smaller $K_{0}$ is preferred to support the astrophysical data of $\Lambda_{1.4}$.

\begin{figure*}[ht!]
  \plottwo{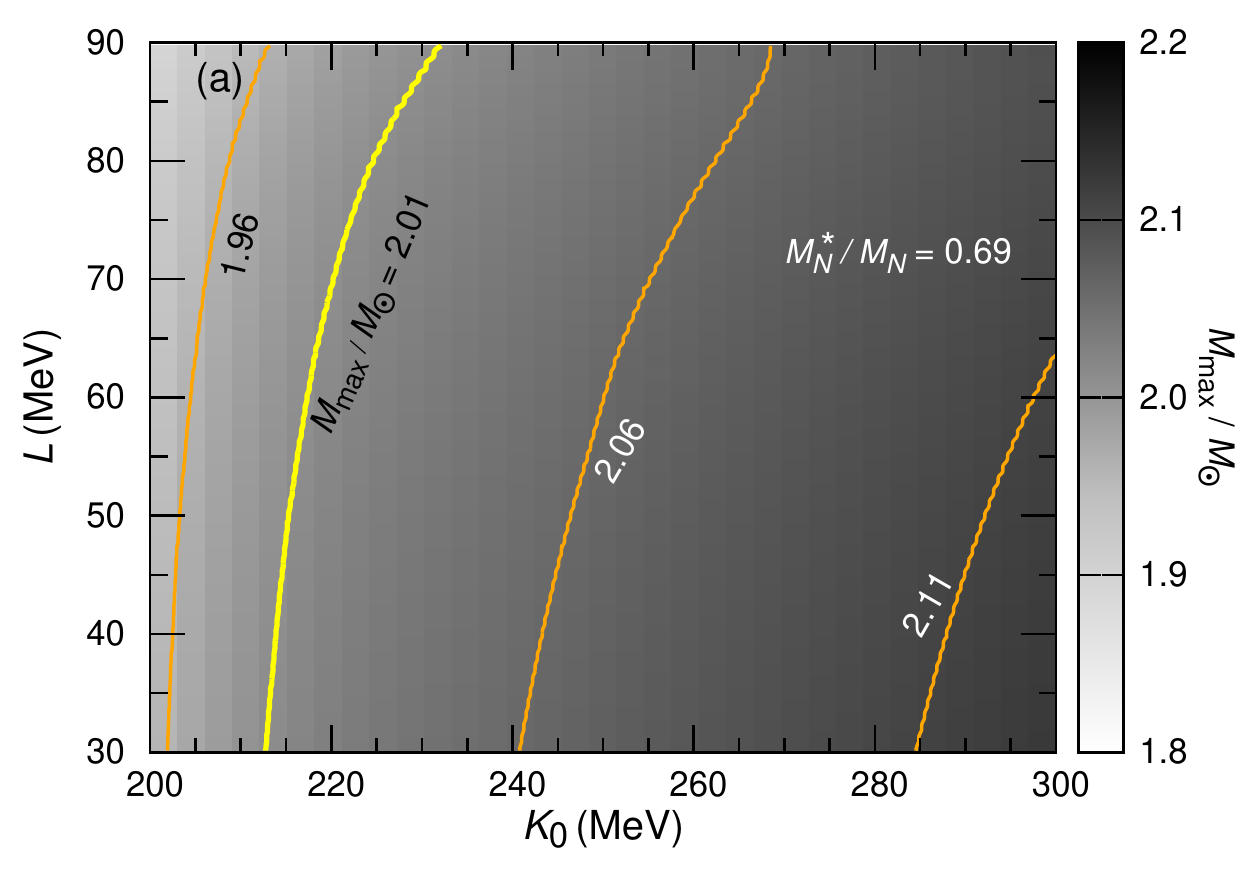}{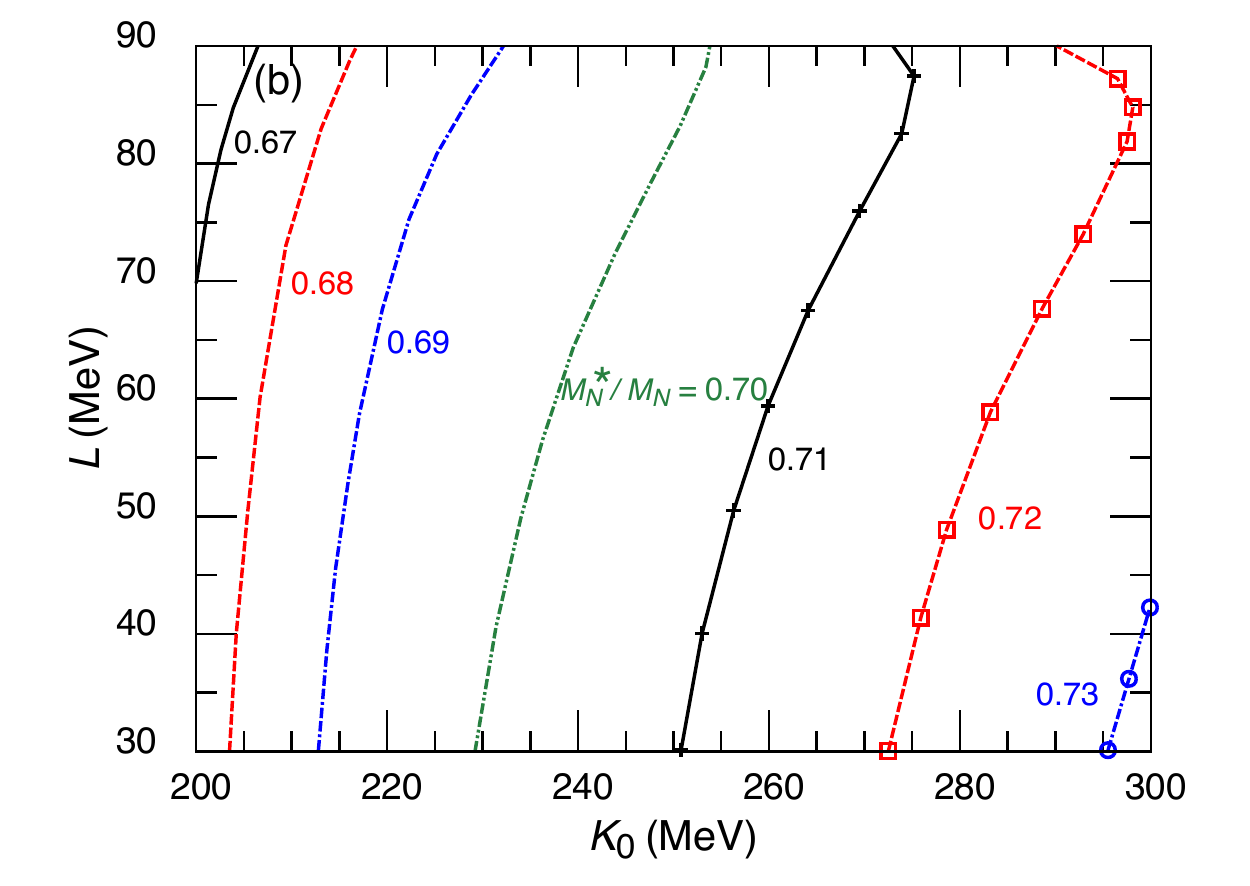}
  \caption{Schematic representation of $M_{\rm max}$ in the $K_{0}$--$L$ plane.
    We show (a) $M_{\rm max}/M_{\sun}$ for $M_{N}^{\ast}/M_{N}=0.69$, and (b) $M_{N}^{\ast}/M_{N}$ for $M_{\rm max}/M_{\sun}=2.01$.
    For details, see the text.
    \label{fig:5}}
\end{figure*}
The contour lines of $M_{\rm max}$ in the $K_{0}$--$L$ plane are presented in Figure \ref{fig:5}.
In Figure \ref{fig:5}(a), we show several lines of $M_{\rm max}/M_{\sun}$ in the case of $M_{N}^{\ast}/M_{N} = 0.69$, which is the favorable condition for supporting $U_{N}^{\rm SEP}$ as already explained in Figure \ref{fig:1}.
It is found that $M_{\rm max}$ is sensitive to $K_{0}$, while $L$ has a little influence on $M_{\rm max}$.
The larger $K_{0}$ and the smaller $L$ are required to support the heavier $M_{\rm max}$, and thus the preferable correlation between $K_{0}$ and $L$ is restricted in the right-hand region above the yellow line of $M_{\rm max}/M_{\sun}=2.01$ to explain the observed mass of PSR J0348+0432 \citep{Antoniadis:2013pzd}.
In addition, the effect of $M_{N}^{\ast}/M_{N}$ at $n_{0}$ on $M_{\rm max}/M_{\sun}$ is depicted in Figure \ref{fig:5}(b).
We find that the smaller $M_{N}^{\ast}$ can easily satisfy the $2M_{\sun}$ constraint with the smaller $K_{0}$, while the larger $K_{0}$ is needed as $M_{N}^{\ast}$ increases.
We here emphasize that owing to the softness of the EoSs for neutron stars with hyperons in the core, it is possible to impose a stringent constraint on the correlation between $K_{0}$ and $L$.

\begin{figure*}[t!]
  \plottwo{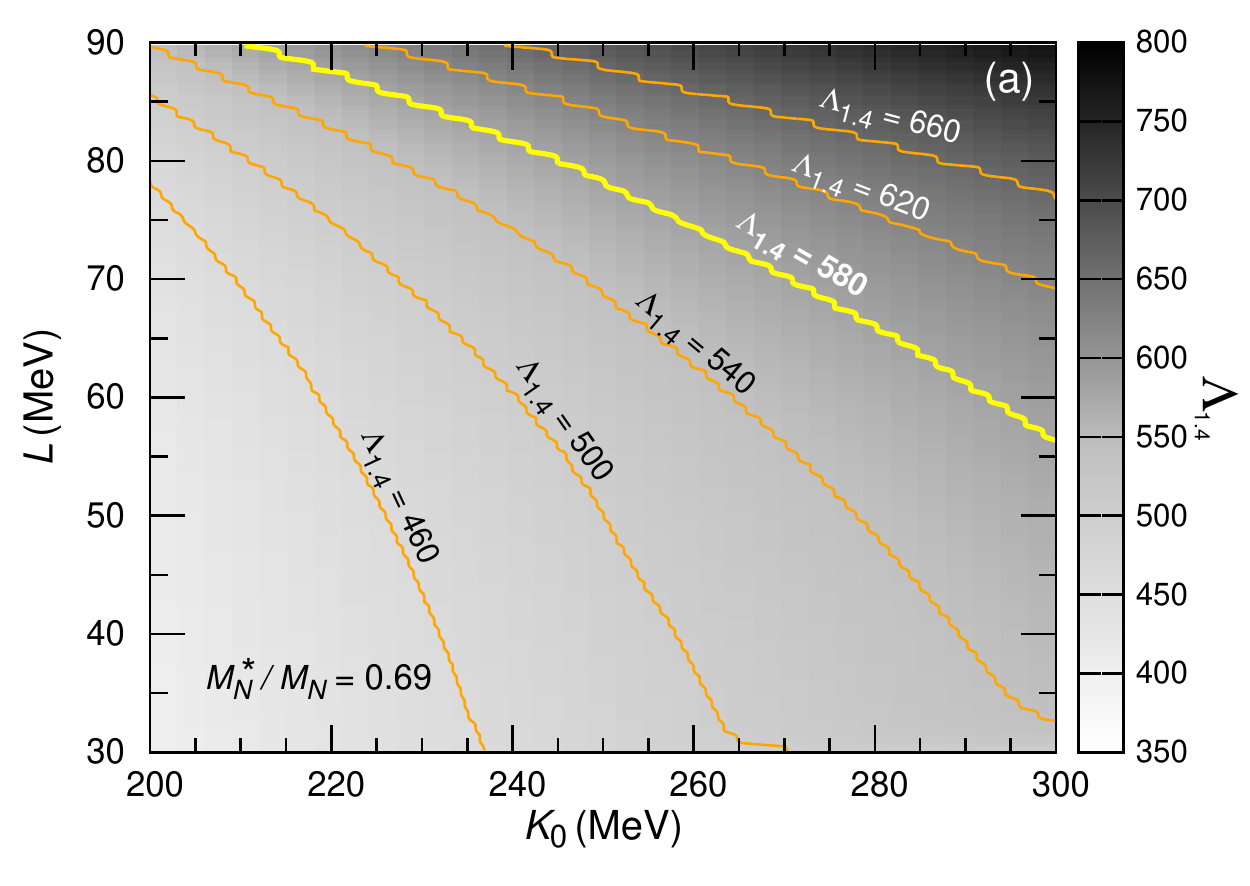}{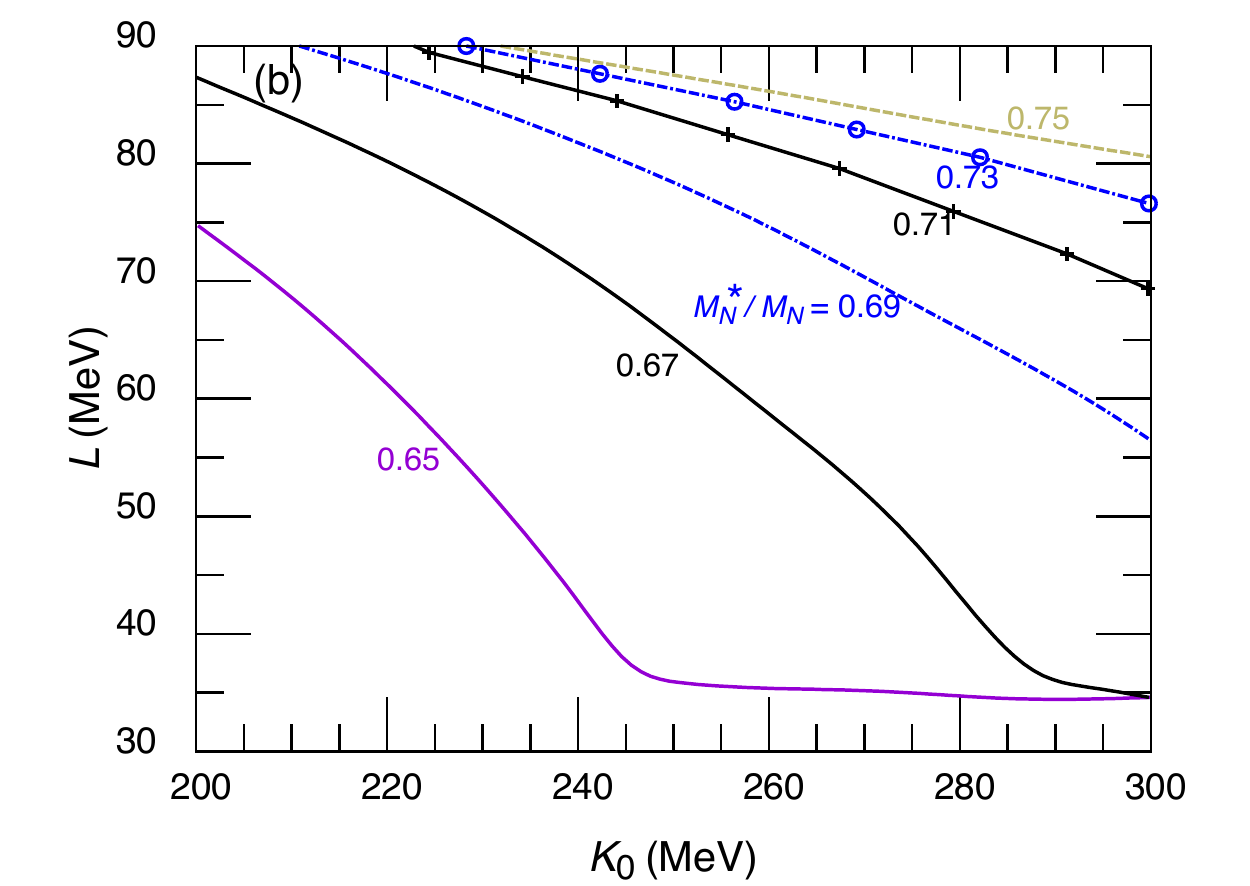}
  \caption{Same as Figure \ref{fig:5}, but for $\Lambda_{1.4}$.
    We show (a) $\Lambda_{1.4}$ for $M_{N}^{\ast}/M_{N}=0.69$, and (b) $M_{N}^{\ast}/M_{N}$ for $\Lambda_{1.4}=580$.
    For details, see the text.
    \label{fig:6}}
\end{figure*}
In Figure \ref{fig:6}, we also present the schematic representation of $\Lambda_{1.4}$ in the $K_{0}$--$L$ plane.
Each line indicates the specific values of $\Lambda_{1.4}$ for $M_{N}^{\ast}/M_{N}=0.69$ in Figure \ref{fig:6}(a).
It is found that $\Lambda_{1.4}$ is quite sensitive to $K_{0}$ and $L$ and, for example, the larger $K_{0}$ and $L$ lead to the larger $\Lambda_{1.4}$.
According to the upper limit based on the GW signals, $\Lambda_{1.4}=580$ \citep{Abbott:2018exr}, the region below the yellow line can be permitted.
Moreover, in Figure \ref{fig:6}(b), we consider the influence of $M_{N}^{\ast}/M_{N}$ on $\Lambda_{1.4}$ in the $K_{0}$--$L$ plane, focusing the upper boundary of $\Lambda_{1.4}$.
As $M_{N}^{\ast}$ decreases, $L$ becomes sensitive to $\Lambda_{1.4}$ compared to $K_{0}$.
We note that, contrary to the case of $M_{\rm max}$ shown in Figure \ref{fig:5}, the existence of hyperons in the core of a neutron star does not affect $\Lambda_{1.4}$ because hyperons do not have any influence on the radii of neutron stars as shown in Figure \ref{fig:3}.

\subsection{Constraints on nuclear saturation properties}
\label{subsec:matter-constraint}

\begin{figure*}[ht!]
  \plottwo{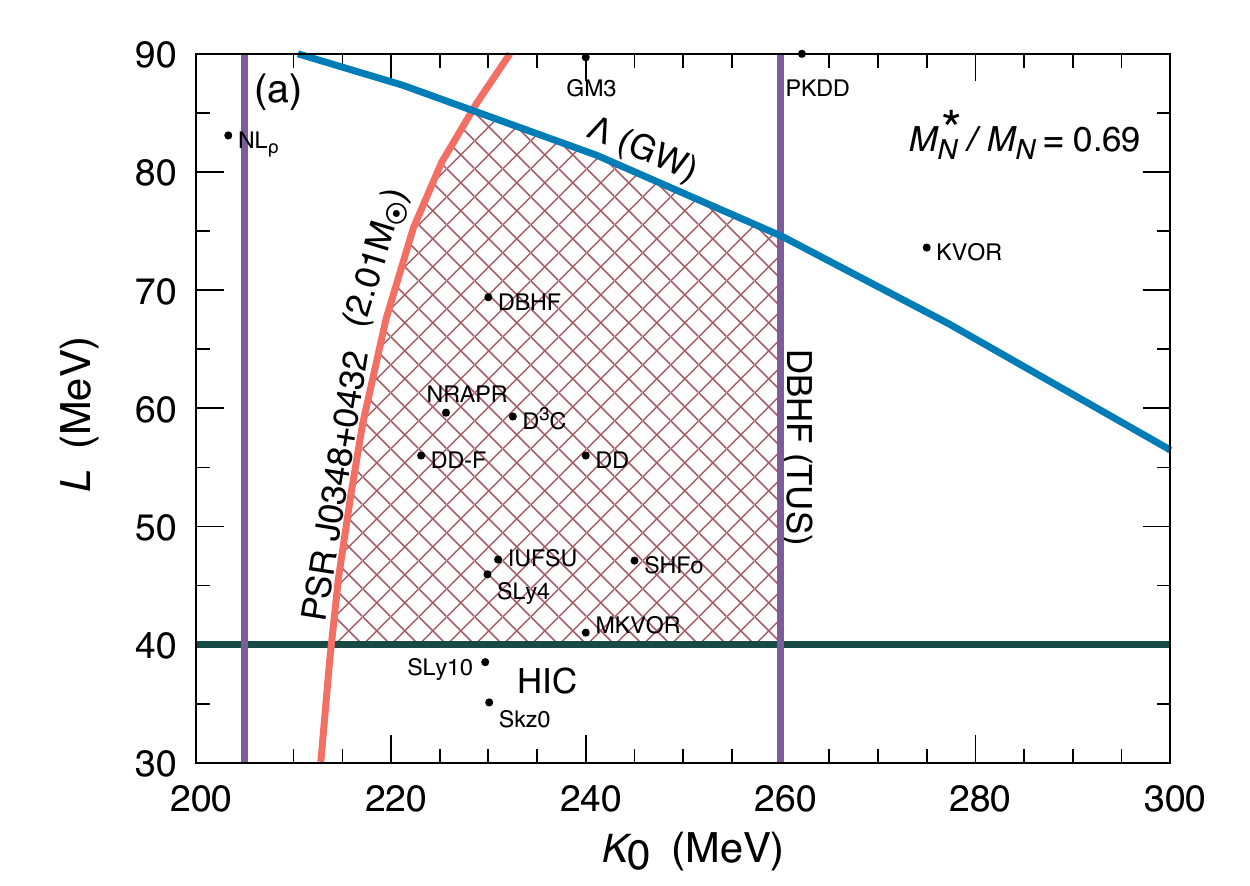}{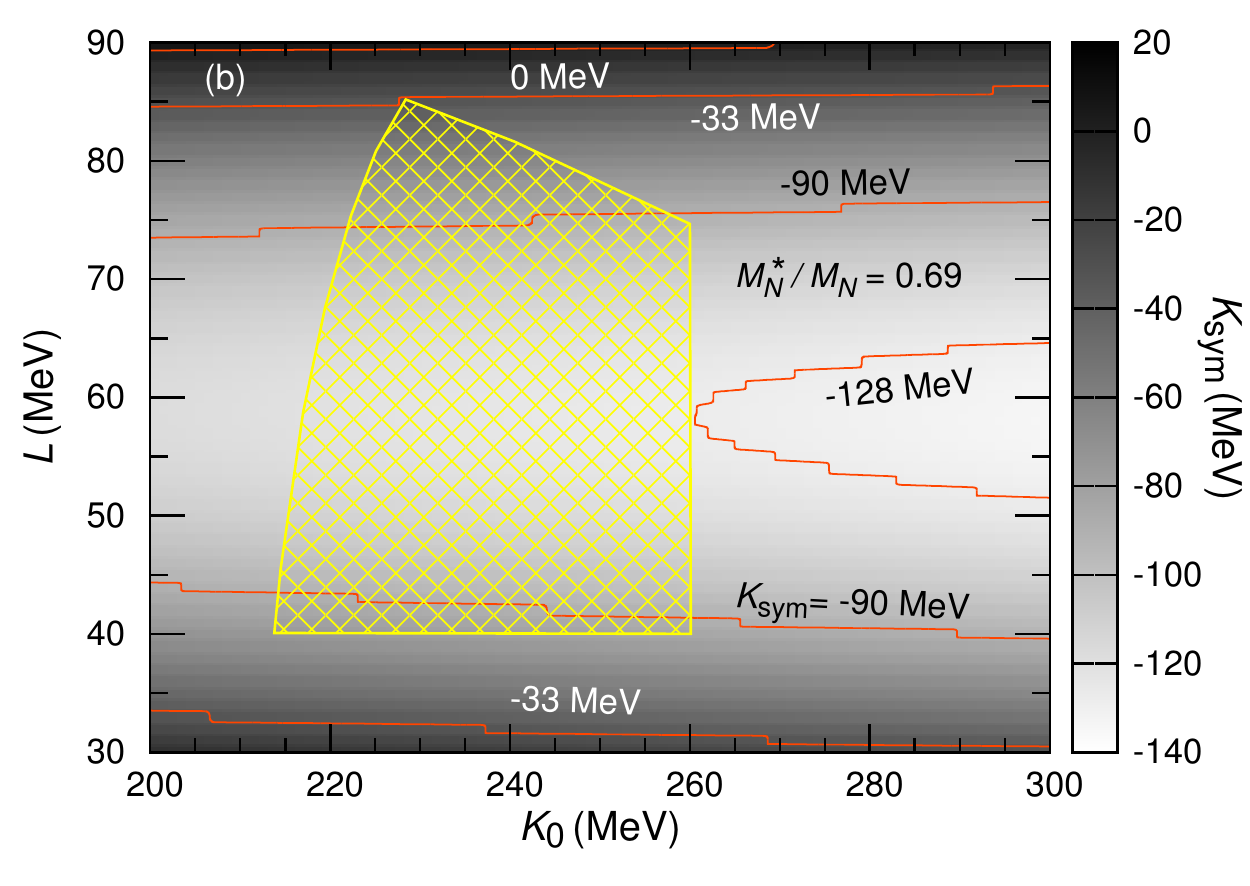}
  \caption{(a) Allowed parameter region in the $K_{0}$--$L$ plane, and (b) constraint on $K_{\rm sym}$ for $M_{N}^{\ast}/M_{N}=0.69$ in the case of $M_{\rm max}/M_{\sun}=2.01$.
    We also show some results calculated by realistic nuclear models in the panel (a) \citep{Dutra:2012mb,Dutra:2014qga,Kolomeitsev:2016sjl}.
    \label{fig:7}}
\end{figure*}
By combining the results obtained from the calculations of nuclear and neutron-star matter, we put restrictions on the nuclear properties at $n_{0}$.
In Figure \ref{fig:7}(a), we show the constrained parameter region in the $K_{0}$--$L$ plane for $M_{N}^{\ast}/M_{N}=0.69$, which is the best fitted value for reproducing $U_{N}^{\rm SEP}$ at $n_{0}$.
As explained in Figure \ref{fig:2}, the lowest limit of $L$ is presented by the data of HIC analysis based on terrestrial experiments, and $K_{0}$ is theoretically constrained by the DBHF (TUS) calculations.
Moreover, we see that $M_{\rm max}$ of PSR J0348+0432 and $\Lambda$ of GW signals, which are based on astrophysical observations, are very useful to give constraints on the relations between $K_{0}$ and $L$.
In addition, it is possible to restrict $K_{\rm sym}$ using the closed, meshed region of $K_{0}$ and $L$ in Figure \ref{fig:7}(b).
It is found that the satisfied ranges of $K_{0}$ and $L$ can be respectively estimated to be $215\le K_{0}$ (MeV) $\le260$ and $40\le L$ (MeV) $\le85$, and the corresponding value of $K_{\rm sym}$ roughly lies in the range of $-128\le K_{\rm sym}$ (MeV) $\le-33$, which is more severe than that of extensive surveys of over 520 theoretical predictions, $-400\le K_{\rm sym}$ (MeV) $\le100$ \citep{Dutra:2012mb,Dutra:2014qga,Li:2020ass}.
The present result is almost the same as the constraints calculated by \cite{Malik:2018zcf}, $-113\le K_{\rm sym}$ (MeV) $\le-52$ and $-141\le K_{\rm sym}$ (MeV) $\le16$.

\begin{figure*}[t!]
  \plottwo{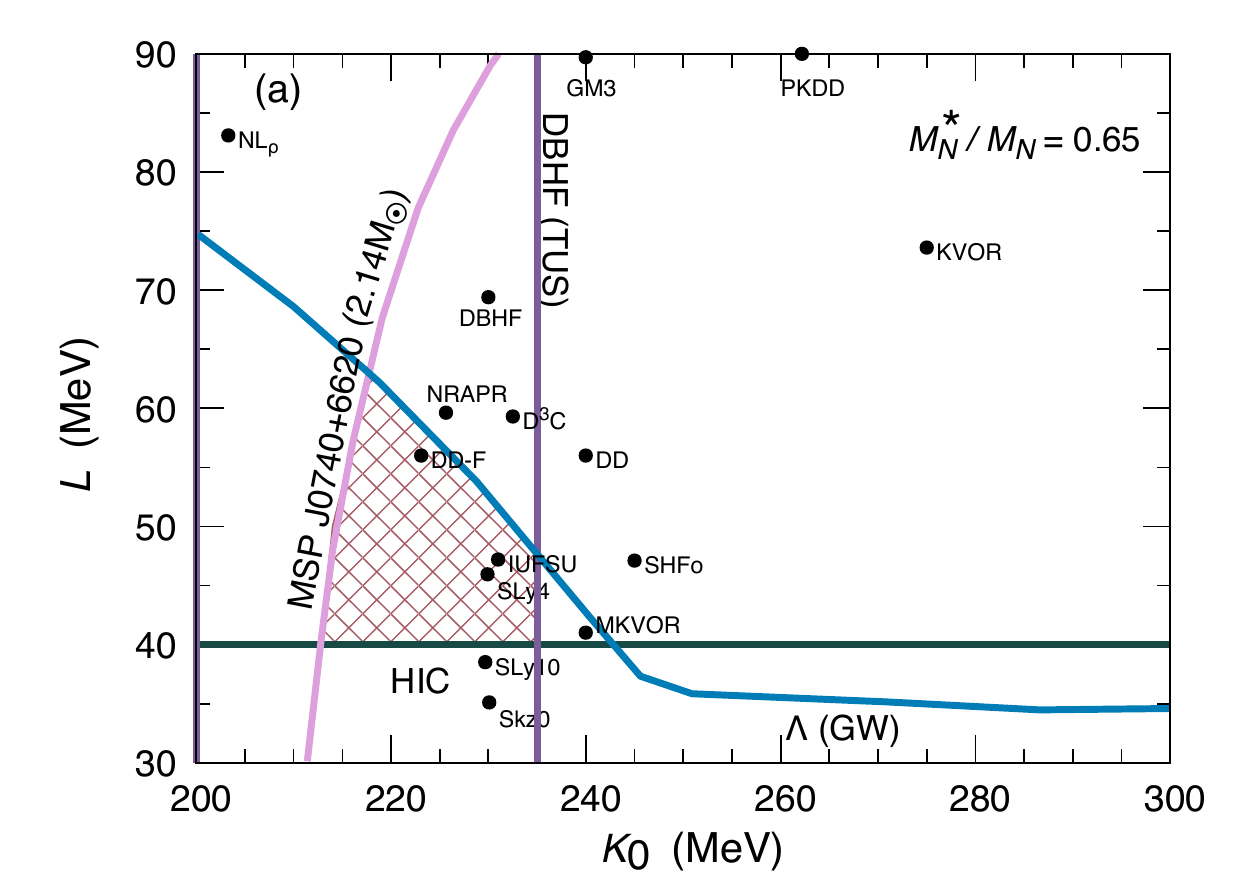}{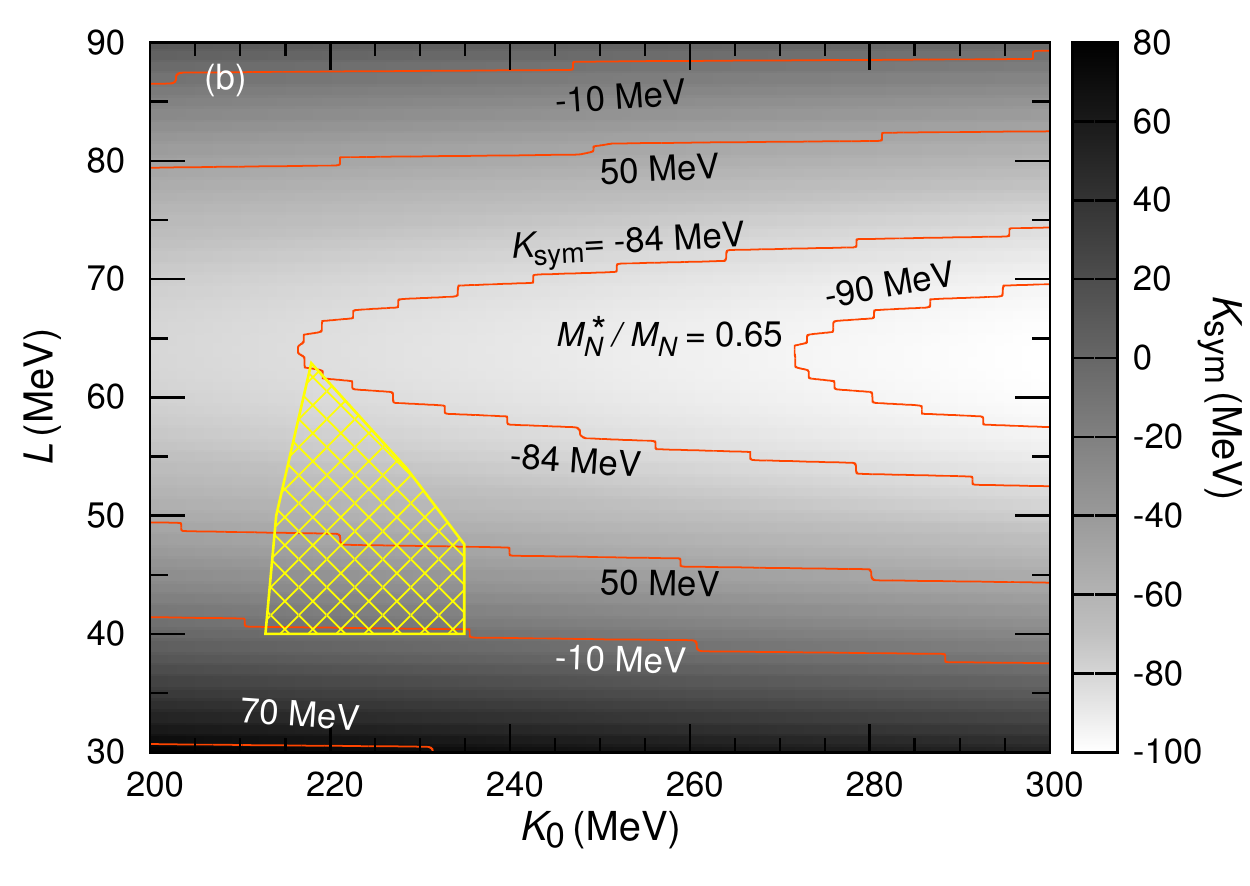}
  \caption{Same as Figure \ref{fig:7}, but for $M_{N}^{\ast}/M_{N}=0.65$ in the case of $M_{\rm max}/M_{\sun}=2.14$.}
  \label{fig:8}
\end{figure*}
Figure \ref{fig:8} shows another acceptable parameter region using the more massive neutron-star condition, $M_{\rm max}/M_{\sun}=2.14$, for $M_{N}^{\ast}/M_{N}=0.65$.
If the heavier mass of a observed neutron star, MSP J0740+6620, is taken into account, then the parameter region is further restricted, and $K_{0}$ and $L$ are respectively estimated to be $215\le K_{0}$ (MeV) $\le235$ and $40\le L$ (MeV) $\le65$ in Figure \ref{fig:8}(a).
Consequently, the more stringent constraint on $K_{\rm sym}$ is given to be $-84\le K_{\rm sym}$ (MeV) $\le-10$ in Figure \ref{fig:8}(b).

\section{Summary}
\label{sec:summary}

  We have studied the properties of nuclear and neutron-star matter using the RMF model with a nonlinear potential.
  In order to restrict the EoS for isospin-asymmetric nuclear matter in the extensive density region, the terrestrial experiments and the recent astrophysical observations of neutron stars and GW signals have been taken into account.
  Moreover, we have investigated the effects of important physical quantities, namely $M_{N}^{\ast}$, $K_{0}$, and $L$, which are still unknown even at $n_{0}$.

  As for the analyses of nuclear properties in the density region below $n_{B}=0.5$ fm$^{-3}$, we have employed the data based on some nuclear experiments and theoretical results.
  We have found that $M_{N}^{\ast}$ at $n_{0}$ is roughly estimated to be $0.65\leq M_{N}^{\ast}/M_{N}\leq0.75$ so as to reproduce $U_{N}^{\rm SEP}$ obtained from the nucleon-nucleus and elastic proton-nucleus scattering data \citep{Li:2013ck,Hama:1990vr}.
  It has been also found that $E_{\rm sym}$ strongly depends on $L$ below $n_{0}$, and $L$ should be larger than 40 MeV to satisfy the experimental results obtained from heavy-ion collisions \citep{Tsang:2012se,Russotto:2016ucm}.
  In addition, compared with the realistic calculations based on the DBHF theory and the $\chi$EFT \citep{Katayama:2013zya,Sammarruca:2014zia}, $K_{0}$ can be restricted to be $180\leq K_{0}$ (MeV) $\leq230$, $210\leq K_{0}$ (MeV) $\leq270$, and $235\leq K_{0}$ (MeV) $\leq305$ in the cases of $M_{N}^{\ast}/M_{N}=0.65$, $0.70$, and $0.75$, respectively.

  Concerning the neutron-star calculations, we have adopted the astrophysical constraints on $M_{\rm max}$ and $\Lambda_{1.4}$ \citep{Antoniadis:2013pzd,Cromartie:2019kug,Abbott:2018exr,Abbott:2018wiz}.
  We have also considered the existence of hyperons in the core to restrict the realistic EoS for neutron stars using SU(3) flavor symmetry \citep{Miyatsu:2011bc,Miyatsu:2013yta,Katayama:2012ge,Weissenborn:2011ut}.
  It has been found that $M_{\rm max}$ is very sensitive to $M_{N}^{\ast}$ at $n_{0}$, and the neutron-star EoSs in the cases of $M_{N}^{\ast}/M_{N}>0.70$ are ruled out to support $2M_{\sun}$ neutron stars with hyperons.
  Additionally, we have found a strong correlation between $K_{0}$ and $\Lambda$, and the smaller $K_{0}$ is preferred to satisfy the astrophysical data of $\Lambda_{1.4}$.

  At last, by combining both calculations of nuclear and neutron-star matter, we have presented the constrained relations between $K_{0}$ and $L$ in Figures \ref{fig:7} and \ref{fig:8}.
  It has been found that, in the case of $M_{N}^{\ast}/M_{N}=0.69$ and $M_{\rm max}/M_{\sun}=2.01$, $K_{0}$ and $L$ can be respectively estimated to be $215\le K_{0}$ (MeV) $\le260$ and $40\le L$ (MeV) $\le85$, and the corresponding value of $K_{\rm sym}$ roughly lies in the range of $-128\le K_{\rm sym}$ (MeV) $\le-33$.
  If we consider the higher limit of a neutron-star mass, it can be possible to impose severe constraints on $K_{0}$, $L$, and $K_{\rm sym}$.

  In conclusion, it has been found that the astrophysical information of massive neutron stars and tidal deformabilities as well as the terrestrial nuclear experimental data plays an important role to restrict the EoS for neutron stars.
  Especially, the softness of the nuclear EoS due to the existence of hyperons in the core gives stringent constraints on the physical quantities, $K_{0}$, $L$, and $K_{\rm sym}$.
  Since the other exotic degrees of freedom in the core of a neutron star and/or the phase transition from hadrons to quarks may influence $M_{\rm max}$ and $\Lambda_{1.4}$ \citep{Alford:2013aca,Miyatsu:2015kwa,Han:2018mtj}, we have to include their effects as well as hyperons.
  We leave them for the future works.

\acknowledgments
This work was supported by JSPS KAKENHI Grant Number JP17K14298 and by the National Research Foundation of Korea (Grant Nos. NRF-2020R1A2C3006177 and NRF-2013M7A1A1075764).

\bibliography{GWNstar}{}

\begin{thebibliography}{}
\expandafter\ifx\csname natexlab\endcsname\relax\def\natexlab#1{#1}\fi
\providecommand{\url}[1]{\href{#1}{#1}}
\providecommand{\dodoi}[1]{doi:~\href{http://doi.org/#1}{\nolinkurl{#1}}}
\providecommand{\doeprint}[1]{\href{http://ascl.net/#1}{\nolinkurl{http://ascl.net/#1}}}
\providecommand{\doarXiv}[1]{\href{https://arxiv.org/abs/#1}{\nolinkurl{https://arxiv.org/abs/#1}}}

\bibitem[{Abbott {et~al.}(2018)}]{Abbott:2018exr}
Abbott, B., {et~al.} 2018, Phys. Rev. Lett., 121, 161101,
  \dodoi{10.1103/PhysRevLett.121.161101}

\bibitem[{Abbott {et~al.}(2019)}]{Abbott:2018wiz}
---. 2019, Phys. Rev. X, 9, 011001, \dodoi{10.1103/PhysRevX.9.011001}

\bibitem[{Alford {et~al.}(2013)Alford, Han, \& Prakash}]{Alford:2013aca}
Alford, M.~G., Han, S., \& Prakash, M. 2013, Phys. Rev. D, 88, 083013,
  \dodoi{10.1103/PhysRevD.88.083013}

\bibitem[{Annala {et~al.}(2018)Annala, Gorda, Kurkela, \&
  Vuorinen}]{Annala:2017llu}
Annala, E., Gorda, T., Kurkela, A., \& Vuorinen, A. 2018, Phys. Rev. Lett.,
  120, 172703, \dodoi{10.1103/PhysRevLett.120.172703}

\bibitem[{Antoniadis {et~al.}(2013)}]{Antoniadis:2013pzd}
Antoniadis, J., {et~al.} 2013, Science, 340, 6131,
  \dodoi{10.1126/science.1233232}

\bibitem[{Arzoumanian {et~al.}(2018)}]{Arzoumanian:2017puf}
Arzoumanian, Z., {et~al.} 2018, Astrophys. J. Suppl., 235, 37,
  \dodoi{10.3847/1538-4365/aab5b0}

\bibitem[{Baldo \& Burgio(2016)}]{Baldo:2016jhp}
Baldo, M., \& Burgio, G.~F. 2016, Prog. Part. Nucl. Phys., 91, 203,
  \dodoi{10.1016/j.ppnp.2016.06.006}

\bibitem[{Boguta \& Bodmer(1977)}]{Boguta:1977xi}
Boguta, J., \& Bodmer, A.~R. 1977, Nucl. Phys., A292, 413,
  \dodoi{10.1016/0375-9474(77)90626-1}

\bibitem[{Capano {et~al.}(2020)Capano, Tews, Brown, Margalit, De, Kumar, Brown,
  Krishnan, \& Reddy}]{Capano:2019eae}
Capano, C.~D., Tews, I., Brown, S.~M., {et~al.} 2020, Nature Astron., 4, 625,
  \dodoi{10.1038/s41550-020-1014-6}

\bibitem[{Chatziioannou {et~al.}(2018)Chatziioannou, Haster, \&
  Zimmerman}]{Chatziioannou:2018vzf}
Chatziioannou, K., Haster, C.-J., \& Zimmerman, A. 2018, Phys. Rev. D, 97,
  104036, \dodoi{10.1103/PhysRevD.97.104036}

\bibitem[{Chen {et~al.}(2009)Chen, Cai, Ko, Li, Shen, \& Xu}]{Chen:2009wv}
Chen, L.-W., Cai, B.-J., Ko, C.~M., {et~al.} 2009, Phys. Rev., C80, 014322,
  \dodoi{10.1103/PhysRevC.80.014322}

\bibitem[{Chen {et~al.}(2007)Chen, Ko, \& Li}]{Chen:2007ih}
Chen, L.-W., Ko, C.~M., \& Li, B.-A. 2007, Phys. Rev., C76, 054316,
  \dodoi{10.1103/PhysRevC.76.054316}

\bibitem[{Cromartie {et~al.}(2019)}]{Cromartie:2019kug}
Cromartie, H.~T., {et~al.} 2019, Nature Astron., 4, 72,
  \dodoi{10.1038/s41550-019-0880-2}

\bibitem[{Danielewicz \& Lee(2009)}]{Danielewicz:2008cm}
Danielewicz, P., \& Lee, J. 2009, Nucl. Phys. A, 818, 36,
  \dodoi{10.1016/j.nuclphysa.2008.11.007}

\bibitem[{De {et~al.}(2018)De, Finstad, Lattimer, Brown, Berger, \&
  Biwer}]{De:2018uhw}
De, S., Finstad, D., Lattimer, J.~M., {et~al.} 2018, Phys. Rev. Lett., 121,
  091102, \dodoi{10.1103/PhysRevLett.121.091102}

\bibitem[{Demorest {et~al.}(2010)Demorest, Pennucci, Ransom, Roberts, \&
  Hessels}]{Demorest:2010bx}
Demorest, P., Pennucci, T., Ransom, S., Roberts, M., \& Hessels, J. 2010,
  Nature, 467, 1081, \dodoi{10.1038/nature09466}

\bibitem[{Dutra {et~al.}(2012)Dutra, Lourenco, {Sa Martins}, Delfino, Stone, \&
  Stevenson}]{Dutra:2012mb}
Dutra, M., Lourenco, O., {Sa Martins}, J.~S., {et~al.} 2012, Phys. Rev., C85,
  035201, \dodoi{10.1103/PhysRevC.85.035201}

\bibitem[{Dutra {et~al.}(2014)Dutra, Louren{\c c}o, Avancini, Carlson, Delfino,
  Menezes, Provid{\^e}ncia, Typel, \& Stone}]{Dutra:2014qga}
Dutra, M., Louren{\c c}o, O., Avancini, S.~S., {et~al.} 2014, Phys. Rev., C90,
  055203, \dodoi{10.1103/PhysRevC.90.055203}

\bibitem[{Fattoyev {et~al.}(2010)Fattoyev, Horowitz, Piekarewicz, \&
  Shen}]{Fattoyev:2010mx}
Fattoyev, F., Horowitz, C., Piekarewicz, J., \& Shen, G. 2010, Phys. Rev. C,
  82, 055803, \dodoi{10.1103/PhysRevC.82.055803}

\bibitem[{Fattoyev {et~al.}(2018)Fattoyev, Piekarewicz, \&
  Horowitz}]{Fattoyev:2017jql}
Fattoyev, F.~J., Piekarewicz, J., \& Horowitz, C.~J. 2018, Phys. Rev. Lett.,
  120, 172702, \dodoi{10.1103/PhysRevLett.120.172702}

\bibitem[{Fortin {et~al.}(2020)Fortin, Raduta, Avancini, \&
  Provid{\^e}ncia}]{Fortin:2020qin}
Fortin, M., Raduta, A.~R., Avancini, S., \& Provid{\^e}ncia, C. 2020, Phys.
  Rev. D, 101, 034017, \dodoi{10.1103/PhysRevD.101.034017}

\bibitem[{Glendenning(1997)}]{Glendenning:1997wn}
Glendenning, N. 1997, {Compact stars: Nuclear physics, particle physics, and
  general relativity}

\bibitem[{Glendenning \& Moszkowski(1991)}]{Glendenning:1991es}
Glendenning, N., \& Moszkowski, S. 1991, Phys. Rev. Lett., 67, 2414,
  \dodoi{10.1103/PhysRevLett.67.2414}

\bibitem[{Hama {et~al.}(1990)Hama, Clark, Cooper, Sherif, \&
  Mercer}]{Hama:1990vr}
Hama, S., Clark, B.~C., Cooper, E.~D., Sherif, H.~S., \& Mercer, R.~L. 1990,
  Phys. Rev., C41, 2737, \dodoi{10.1103/PhysRevC.41.2737}

\bibitem[{Han \& Steiner(2019)}]{Han:2018mtj}
Han, S., \& Steiner, A.~W. 2019, Phys. Rev. D, 99, 083014,
  \dodoi{10.1103/PhysRevD.99.083014}

\bibitem[{Hewish {et~al.}(1968)Hewish, Bell, Pilkington, Scott, \&
  Collins}]{Hewish:1968bj}
Hewish, A., Bell, S., Pilkington, J., Scott, P., \& Collins, R. 1968, Nature,
  217, 709, \dodoi{10.1038/217709a0}

\bibitem[{{Hewish} \& {Okoye}(1965)}]{1965Natur.207...59H}
{Hewish}, A., \& {Okoye}, S.~E. 1965, \nat, 207, 59, \dodoi{10.1038/207059a0}

\bibitem[{Hinderer(2008)}]{Hinderer:2007mb}
Hinderer, T. 2008, Astrophys. J., 677, 1216, \dodoi{10.1086/533487}

\bibitem[{Hinderer {et~al.}(2010)Hinderer, Lackey, Lang, \&
  Read}]{Hinderer:2009ca}
Hinderer, T., Lackey, B.~D., Lang, R.~N., \& Read, J.~S. 2010, Phys. Rev. D,
  81, 123016, \dodoi{10.1103/PhysRevD.81.123016}

\bibitem[{Hornick {et~al.}(2018)Hornick, Tolos, Zacchi, Christian, \&
  Schaffner-Bielich}]{Hornick:2018kfi}
Hornick, N., Tolos, L., Zacchi, A., Christian, J.-E., \& Schaffner-Bielich, J.
  2018, Phys. Rev., C98, 065804, \dodoi{10.1103/PhysRevC.98.065804}

\bibitem[{Jaminon {et~al.}(1981)Jaminon, Mahaux, \& Rochus}]{Jaminon:1981xg}
Jaminon, M., Mahaux, C., \& Rochus, P. 1981, Nucl. Phys., A365, 371,
  \dodoi{10.1016/0375-9474(81)90397-3}

\bibitem[{Katayama {et~al.}(2012)Katayama, Miyatsu, \& Saito}]{Katayama:2012ge}
Katayama, T., Miyatsu, T., \& Saito, K. 2012, Astrophys. J. Suppl., 203, 22,
  \dodoi{10.1088/0067-0049/203/2/22}

\bibitem[{Katayama \& Saito(2013)}]{Katayama:2013zya}
Katayama, T., \& Saito, K. 2013, Phys. Rev. C, 88, 035805,
  \dodoi{10.1103/PhysRevC.88.035805}

\bibitem[{Kim {et~al.}(2018)Kim, Lim, Kwak, Hyun, \& Lee}]{Kim:2018aoi}
Kim, Y.-M., Lim, Y., Kwak, K., Hyun, C.~H., \& Lee, C.-H. 2018, Phys. Rev. C,
  98, 065805, \dodoi{10.1103/PhysRevC.98.065805}

\bibitem[{Krastev \& Li(2019)}]{Krastev:2018nwr}
Krastev, P.~G., \& Li, B.-A. 2019, Comments Nucl. Part. Phys., 46, 074001,
  \dodoi{10.1088/1361-6471/ab1a7a}

\bibitem[{Kumar {et~al.}(2017)Kumar, Biswal, \& Patra}]{Kumar:2016dks}
Kumar, B., Biswal, S., \& Patra, S. 2017, Phys. Rev. C, 95, 015801,
  \dodoi{10.1103/PhysRevC.95.015801}

\bibitem[{Lattimer \& Prakash(2007)}]{Lattimer:2006xb}
Lattimer, J.~M., \& Prakash, M. 2007, Phys. Rept., 442, 109,
  \dodoi{10.1016/j.physrep.2007.02.003}

\bibitem[{Li {et~al.}(2008)Li, Chen, \& Ko}]{Li:2008gp}
Li, B.-A., Chen, L.-W., \& Ko, C.~M. 2008, Phys. Rept., 464, 113,
  \dodoi{10.1016/j.physrep.2008.04.005}

\bibitem[{Li \& Han(2013)}]{Li:2013ola}
Li, B.-A., \& Han, X. 2013, Phys. Lett., B727, 276,
  \dodoi{10.1016/j.physletb.2013.10.006}

\bibitem[{Li {et~al.}(2019)Li, Krastev, Wen, \& Zhang}]{Li:2019xxz}
Li, B.-A., Krastev, P.~G., Wen, D.-H., \& Zhang, N.-B. 2019, Eur. Phys. J. A,
  55, 117, \dodoi{10.1140/epja/i2019-12780-8}

\bibitem[{Li \& Magno(2020)}]{Li:2020ass}
Li, B.-A., \& Magno, M. 2020, Phys. Rev. C, 102, 045807,
  \dodoi{10.1103/PhysRevC.102.045807}

\bibitem[{Li {et~al.}(2014)Li, Ramos, Verde, \& Vidana}]{Li:2014oda}
Li, B.-A., Ramos, A., Verde, G., \& Vidana, I. 2014, Eur. Phys. J. A, 50, 9,
  \dodoi{10.1140/epja/i2014-14009-x}

\bibitem[{Li {et~al.}(2018)Li, Yan, Geng, Huang, \& Zong}]{Li:2018ayl}
Li, C.-M., Yan, Y., Geng, J.-J., Huang, Y.-F., \& Zong, H.-S. 2018, Phys. Rev.
  D, 98, 083013, \dodoi{10.1103/PhysRevD.98.083013}

\bibitem[{Li \& Sedrakian(2019)}]{Li:2019tjx}
Li, J.~J., \& Sedrakian, A. 2019, Astrophys. J. Lett., 874, L22,
  \dodoi{10.3847/2041-8213/ab1090}

\bibitem[{Li {et~al.}(2013)Li, Cai, Chen, Chen, Li, \& Xu}]{Li:2013ck}
Li, X.-H., Cai, B.-J., Chen, L.-W., {et~al.} 2013, Phys. Lett., B721, 101,
  \dodoi{10.1016/j.physletb.2013.03.005}

\bibitem[{Lim \& Holt(2018)}]{Lim:2018bkq}
Lim, Y., \& Holt, J.~W. 2018, Phys. Rev. Lett., 121, 062701,
  \dodoi{10.1103/PhysRevLett.121.062701}

\bibitem[{Louren{\c c}o {et~al.}(2019)Louren{\c c}o, Dutra, Lenzi, Flores, \&
  Menezes}]{Lourenco:2018dvh}
Louren{\c c}o, O., Dutra, M., Lenzi, C.~H., Flores, C.~V., \& Menezes, D.~P.
  2019, Phys. Rev. C, 99, 045202, \dodoi{10.1103/PhysRevC.99.045202}

\bibitem[{Malik {et~al.}(2018)Malik, Alam, Fortin, Provid{\^e}ncia, Agrawal,
  Jha, Kumar, \& Patra}]{Malik:2018zcf}
Malik, T., Alam, N., Fortin, M., {et~al.} 2018, Phys. Rev. C, 98, 035804,
  \dodoi{10.1103/PhysRevC.98.035804}

\bibitem[{Miyatsu {et~al.}(2020)Miyatsu, Cheoun, Ishizuka, Kim, Maruyama, \&
  Saito}]{Miyatsu:2020vzi}
Miyatsu, T., Cheoun, M.-K., Ishizuka, C., {et~al.} 2020, Phys. Lett. B, 803,
  135282, \dodoi{10.1016/j.physletb.2020.135282}

\bibitem[{Miyatsu {et~al.}(2013{\natexlab{a}})Miyatsu, Cheoun, \&
  Saito}]{Miyatsu:2013yta}
Miyatsu, T., Cheoun, M.-K., \& Saito, K. 2013{\natexlab{a}}, Phys. Rev., C88,
  015802, \dodoi{10.1103/PhysRevC.88.015802}

\bibitem[{Miyatsu {et~al.}(2015)Miyatsu, Cheoun, \& Saito}]{Miyatsu:2015kwa}
---. 2015, Astrophys. J., 813, 135, \dodoi{10.1088/0004-637X/813/2/135}

\bibitem[{Miyatsu {et~al.}(2012)Miyatsu, Katayama, \& Saito}]{Miyatsu:2011bc}
Miyatsu, T., Katayama, T., \& Saito, K. 2012, Phys. Lett. B, 709, 242,
  \dodoi{10.1016/j.physletb.2012.02.009}

\bibitem[{Miyatsu {et~al.}(2013{\natexlab{b}})Miyatsu, Yamamuro, \&
  Nakazato}]{Miyatsu:2013hea}
Miyatsu, T., Yamamuro, S., \& Nakazato, K. 2013{\natexlab{b}}, Astrophys. J.,
  777, 4, \dodoi{10.1088/0004-637X/777/1/4}

\bibitem[{Most {et~al.}(2018)Most, Weih, Rezzolla, \&
  Schaffner-Bielich}]{Most:2018hfd}
Most, E.~R., Weih, L.~R., Rezzolla, L., \& Schaffner-Bielich, J. 2018, Phys.
  Rev. Lett., 120, 261103, \dodoi{10.1103/PhysRevLett.120.261103}

\bibitem[{Oppenheimer \& Volkoff(1939)}]{Oppenheimer:1939ne}
Oppenheimer, J., \& Volkoff, G. 1939, Phys. Rev., 55, 374,
  \dodoi{10.1103/PhysRev.55.374}

\bibitem[{Paschalidis {et~al.}(2018)Paschalidis, Yagi, Alvarez-Castillo,
  Blaschke, \& Sedrakian}]{Paschalidis:2017qmb}
Paschalidis, V., Yagi, K., Alvarez-Castillo, D., Blaschke, D.~B., \& Sedrakian,
  A. 2018, Phys. Rev. D, 97, 084038, \dodoi{10.1103/PhysRevD.97.084038}

\bibitem[{Radice {et~al.}(2018)Radice, Perego, Zappa, \&
  Bernuzzi}]{Radice:2017lry}
Radice, D., Perego, A., Zappa, F., \& Bernuzzi, S. 2018, Astrophys. J. Lett.,
  852, L29, \dodoi{10.3847/2041-8213/aaa402}

\bibitem[{Raithel {et~al.}(2018)Raithel, {\"O}zel, \&
  Psaltis}]{Raithel:2018ncd}
Raithel, C., {\"O}zel, F., \& Psaltis, D. 2018, Astrophys. J. Lett., 857, L23,
  \dodoi{10.3847/2041-8213/aabcbf}

\bibitem[{Raithel \& Ozel(2019)}]{Raithel:2019ejc}
Raithel, C.~A., \& Ozel, F. 2019, Astrophys. J., 885, 121,
  \dodoi{10.3847/1538-4357/ab48e6}

\bibitem[{Ribes {et~al.}(2019)Ribes, Ramos, Tolos, Gonzalez-Boquera, \&
  Centelles}]{Ribes:2019kno}
Ribes, P., Ramos, A., Tolos, L., Gonzalez-Boquera, C., \& Centelles, M. 2019,
  Astrophys. J., 883, 168, \dodoi{10.3847/1538-4357/ab3a93}

\bibitem[{Rijken {et~al.}(2010)Rijken, Nagels, \& Yamamoto}]{Rijken:2010zzb}
Rijken, T.~A., Nagels, M.~M., \& Yamamoto, Y. 2010, Prog. Theor. Phys. Suppl.,
  185, 14, \dodoi{10.1143/PTPS.185.14}

\bibitem[{Russotto {et~al.}(2016)}]{Russotto:2016ucm}
Russotto, P., {et~al.} 2016, Phys. Rev. C, 94, 034608,
  \dodoi{10.1103/PhysRevC.94.034608}

\bibitem[{Sahoo {et~al.}(2019)Sahoo, Mishra, Mohanty, Panda, \&
  Barik}]{Sahoo:2019qaq}
Sahoo, H.~S., Mishra, R., Mohanty, D.~K., Panda, P.~K., \& Barik, N. 2019,
  Phys. Rev. C, 99, 055803, \dodoi{10.1103/PhysRevC.99.055803}

\bibitem[{Sammarruca {et~al.}(2015)Sammarruca, Coraggio, Holt, Itaco,
  Machleidt, \& Marcucci}]{Sammarruca:2014zia}
Sammarruca, F., Coraggio, L., Holt, J., {et~al.} 2015, Phys. Rev. C, 91,
  054311, \dodoi{10.1103/PhysRevC.91.054311}

\bibitem[{Schaffner \& Mishustin(1996)}]{Schaffner:1995th}
Schaffner, J., \& Mishustin, I.~N. 1996, Phys. Rev. C, 53, 1416,
  \dodoi{10.1103/PhysRevC.53.1416}

\bibitem[{Schaffner-Bielich(2008)}]{SchaffnerBielich:2008kb}
Schaffner-Bielich, J. 2008, Nucl. Phys. A, 804, 309,
  \dodoi{10.1016/j.nuclphysa.2008.01.005}

\bibitem[{Serot \& Walecka(1986)}]{Serot:1984ey}
Serot, B.~D., \& Walecka, J.~D. 1986, Adv. Nucl. Phys., 16, 1

\bibitem[{Takahashi {et~al.}(2001)}]{Takahashi:2001nm}
Takahashi, H., {et~al.} 2001, Phys. Rev. Lett., 87, 212502,
  \dodoi{10.1103/PhysRevLett.87.212502}

\bibitem[{Tews {et~al.}(2017)Tews, Lattimer, Ohnishi, \&
  Kolomeitsev}]{Kolomeitsev:2016sjl}
Tews, I., Lattimer, J.~M., Ohnishi, A., \& Kolomeitsev, E.~E. 2017, Astrophys.
  J., 848, 105, \dodoi{10.3847/1538-4357/aa8db9}

\bibitem[{Tews {et~al.}(2018)Tews, Margueron, \& Reddy}]{Tews:2018iwm}
Tews, I., Margueron, J., \& Reddy, S. 2018, Phys. Rev. C, 98, 045804,
  \dodoi{10.1103/PhysRevC.98.045804}

\bibitem[{Tews {et~al.}(2019)Tews, Margueron, \& Reddy}]{Tews:2019cap}
---. 2019, Eur. Phys. J. A, 55, 97, \dodoi{10.1140/epja/i2019-12774-6}

\bibitem[{Todd-Rutel \& Piekarewicz(2005{\natexlab{a}})}]{ToddRutel:2005zz}
Todd-Rutel, B., \& Piekarewicz, J. 2005{\natexlab{a}}, Phys. Rev. Lett., 95,
  122501, \dodoi{10.1103/PhysRevLett.95.122501}

\bibitem[{Todd-Rutel \& Piekarewicz(2005{\natexlab{b}})}]{ToddRutel:2005fa}
Todd-Rutel, B.~G., \& Piekarewicz, J. 2005{\natexlab{b}}, Phys. Rev. Lett., 95,
  122501, \dodoi{10.1103/PhysRevLett.95.122501}

\bibitem[{Tolman(1934)}]{Tolman:1934za}
Tolman, R.~C. 1934, Proc. Nat. Acad. Sci., 20, 169,
  \dodoi{10.1073/pnas.20.3.169}

\bibitem[{Tsang {et~al.}(2012)}]{Tsang:2012se}
Tsang, M., {et~al.} 2012, Phys. Rev. C, 86, 015803,
  \dodoi{10.1103/PhysRevC.86.015803}

\bibitem[{Walecka(1974)}]{Walecka:1974qa}
Walecka, J. 1974, Annals Phys., 83, 491, \dodoi{10.1016/0003-4916(74)90208-5}

\bibitem[{Wei {et~al.}(2019)Wei, Figura, Burgio, Chen, \&
  Schulze}]{Wei:2018dyy}
Wei, J., Figura, A., Burgio, G., Chen, H., \& Schulze, H. 2019, J. Phys. G, 46,
  034001, \dodoi{10.1088/1361-6471/aaf95c}

\bibitem[{Weissenborn {et~al.}(2012)Weissenborn, Chatterjee, \&
  Schaffner-Bielich}]{Weissenborn:2011ut}
Weissenborn, S., Chatterjee, D., \& Schaffner-Bielich, J. 2012, Phys. Rev.,
  C85, 065802, \dodoi{10.1103/PhysRevC.85.065802; 10.1103/PhysRevC.90.019904}

\bibitem[{Xie \& Li(2019)}]{Xie:2019sqb}
Xie, W.-J., \& Li, B.-A. 2019, Astrophys. J., 883, 174,
  \dodoi{10.3847/1538-4357/ab3f37}

\bibitem[{Yang \& Shen(2008)}]{Yang:2008am}
Yang, F., \& Shen, H. 2008, Phys. Rev. C, 77, 025801,
  \dodoi{10.1103/PhysRevC.77.025801}

\bibitem[{Zhang \& Li(2019)}]{Zhang:2018vbw}
Zhang, N.-B., \& Li, B.-A. 2019, J. Phys. G, 46, 014002,
  \dodoi{10.1088/1361-6471/aaef54}

\bibitem[{Zhang {et~al.}(2018)Zhang, Li, \& Xu}]{Zhang:2018vrx}
Zhang, N.-B., Li, B.-A., \& Xu, J. 2018, Astrophys. J., 859, 90,
  \dodoi{10.3847/1538-4357/aac027}

\bibitem[{Zhao \& Lattimer(2018)}]{Zhao:2018nyf}
Zhao, T., \& Lattimer, J.~M. 2018, Phys. Rev. D, 98, 063020,
  \dodoi{10.1103/PhysRevD.98.063020}

\bibitem[{Zhou {et~al.}(2018)Zhou, Zhou, \& Li}]{Zhou:2017pha}
Zhou, E.-P., Zhou, X., \& Li, A. 2018, Phys. Rev. D, 97, 083015,
  \dodoi{10.1103/PhysRevD.97.083015}

\bibitem[{Zhu {et~al.}(2018)Zhu, Zhou, \& Li}]{Zhu:2018ona}
Zhu, Z.-Y., Zhou, E.-P., \& Li, A. 2018, Astrophys. J., 862, 98,
  \dodoi{10.3847/1538-4357/aacc28}

\end{thebibliography}
\bibliographystyle{aasjournal}

\end{document}